\documentclass{elsart}

\usepackage{graphicx}
\usepackage[figuresright]{rotating}

\newcommand{\AmS}{{\protect\the\textfont2
  A\kern-.1667em\lower.5ex\hbox{M}\kern-.125emS}}

\begin{document}

\begin{frontmatter}
\title{Bimodality: a possible experimental signature of the liquid-gas phase
transition of nuclear matter}

\author[LPC]{M. Pichon}, 
\author[LPC]{B. Tamain},
\author[LPC]{R. Bougault}, 
\author[LPC]{F. Gulminelli}, 
\author[LPC]{O. Lopez},
\author[IPNO]{E. Bonnet}, 
\author[IPNO]{B. Borderie}, 
\author[GANIL]{A. Chbihi},
\author[DAPNIA]{R. Dayras}, 
\author[GANIL]{J.D. Frankland},
\author[CNAM]{E. Galichet}, 
\author[IPNL]{D. Guinet},
\author[IPNL]{P. Lautesse},
\author[IPNO]{N. Le Neindre}, 
\author[Bucharest]{M. P\^arlog},
\author[IPNO]{M.F. Rivet},
\author[Quebec]{R. Roy},      
\author[Naples]{E. Rosato}, 
\author[LPC]{E. Vient},
\author[Naples]{M. Vigilante}, 
\author[DAPNIA]{C. Volant}
\author[GANIL]{J.P. Wieleczko},
\author[Pologne]{B. Zwieglinski}

\collaboration{INDRA and ALADIN collaborations}  
       
\address[LPC]{LPC, IN2P3/CNRS, ENSICaen and University, F-14050 Caen cedex,France}
\address[GANIL]{GANIL, DSM-CEA/IN2P3-CNRS, BP 5027,F-14076 Caen cedex 5,France}
\address[IPNO]{IPN, IN2P3/CNRS, BP 1, F-91406 Orsay cedex, France}
\address[DAPNIA]{DAPNIA/SPhN, CEA/SAclay, F-91191 Gif-sur-Yvette cedex, France}
\address[IPNL]{IPN, IN2P3/CNRS and University, F-69622 Villeurbanne cedex, France}
\address[Bucharest]{Nat. Inst. for Phys. and Nucl. Eng., RO-76900 Bucharest-Magurele, Romania}
\address[Naples]{Dip. di Sci. Fis. Sez. INFN, Univ. di Napoli, ''Frederico II'', I-80126 Napoli, Italy}
\address[Quebec]{Laboratoire de Physique Nucléaire, Université Laval, Québec, Canada}
\address[CNAM]{CNAM, Laboratoire des Sciences Nucléaires, F-75003 Paris, France}
\address[Pologne]{Soltan Institute for Nuclear Studies, 00-681 Warsaw, Poland}

\maketitle

\begin{abstract}

We have observed a bimodal behaviour of the distribution of the asymmetry between the charges of
the two heaviest products resulting from the decay of the quasi-projectile released in
binary Xe+Sn and Au+Au collisions from 60 to 100 MeV/u. Event sorting has been achieved
through the transverse energy of light charged particles emitted on the quasi-target side,
thus avoiding artificial correlations between the bimodality signal and the sorting variable. 
Bimodality is observed for intermediate
impact parameters for which the quasi-projectile is identified. A simulation shows that the deexcitation step 
rather than the geometry of the collision appears responsible for the bimodal behaviour. 
The influence of mid-rapidity emission has been verified. 
The two bumps of the bimodal distribution correspond to different excitation energies and similar temperatures.
It is also shown
that it is possible to correlate the bimodality signal with a change in the distribution of the heaviest
fragment charge and a peak in potential energy fluctuations. All together, this set of data is coherent with 
what would be expected in a finite system if the corresponding system in the thermodynamic
limit exhibits a first order phase transition.
\end{abstract}

\end{frontmatter}

\section{Introduction}\label{Introduction}

Many experimental features suggestive of the liquid-gas phase transition of nuclear matter have been 
observed in the Fermi energy regime of nucleus-nucleus
collisions. Among these features are: abnormal partial energy fluctuations\cite{Dagos,Lenein}, 
charge correlations\cite{Bord},
a double peaked distribution of an order parameter ("bimodality")\cite{Tam,Pichon}, 
fluctuation properties of the heaviest fragment size\cite{Botet2,Frankland}, Fisher scaling\cite{Moret},
vaporisation\cite{Bord2}, flattening of the caloric curve\cite{Poch,Natowitz}. In the present paper, we
concentrate on the bimodality signal. 

The discontinuity of the order parameter
at a first order phase transition is expected to be replaced in a finite system by a bimodal
distribution of the order parameter close to the transition point\cite{Binder}. If the order parameter is one 
dimensional\cite{Gulmi2005} and if the transition has a finite latent heat, the bimodality should be observed in 
the canonical ensemble, the transition temperature being defined as that at which the two peaks have the same height 
\cite{Chomaz}. In the fragmentation transition case, the size of the heaviest cluster produced in each event 
is an order parameter in many different models\cite{Botet} including the lattice-gas model
of the liquid-gas phase transition\cite{Gulmi2005}. In this picture, when the
size of the heaviest fragment is large, the system is mostly on the liquid-like side, whereas if it is
small one is mostly dealing with a gas-like behaviour. 

\begin{figure}[htb]
\begin{minipage}[t]{65mm}
\begin{center}
\includegraphics[width=6.5cm]{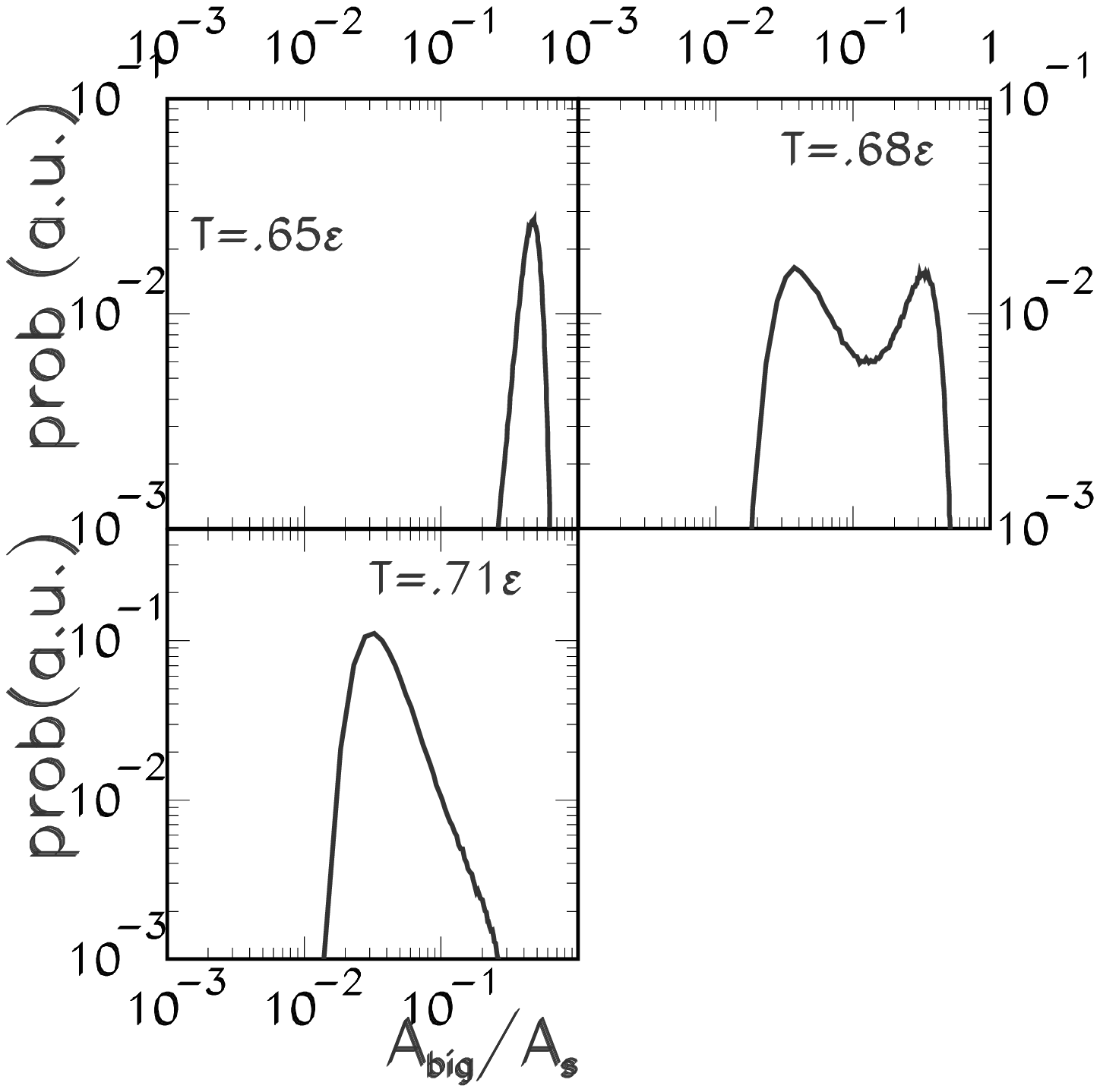}
\end{center}
\vspace{-0cm}
\caption{Size distribution of the largest product obtained with the lattice-gas model
for an equilibrated system of 216 particles in the canonical isobar ensemble at subcritical pressure\cite{Chomaz}. 
$\epsilon$ represents the lattice coupling.
The three plots correspond to temperatures respectively 
below, at and above the transition one.}

\label{fig:gulmi}

\end{minipage}
\hspace{\fill}
\begin{minipage}[t]{65mm}
\begin{center}
\includegraphics[width=6.5cm]{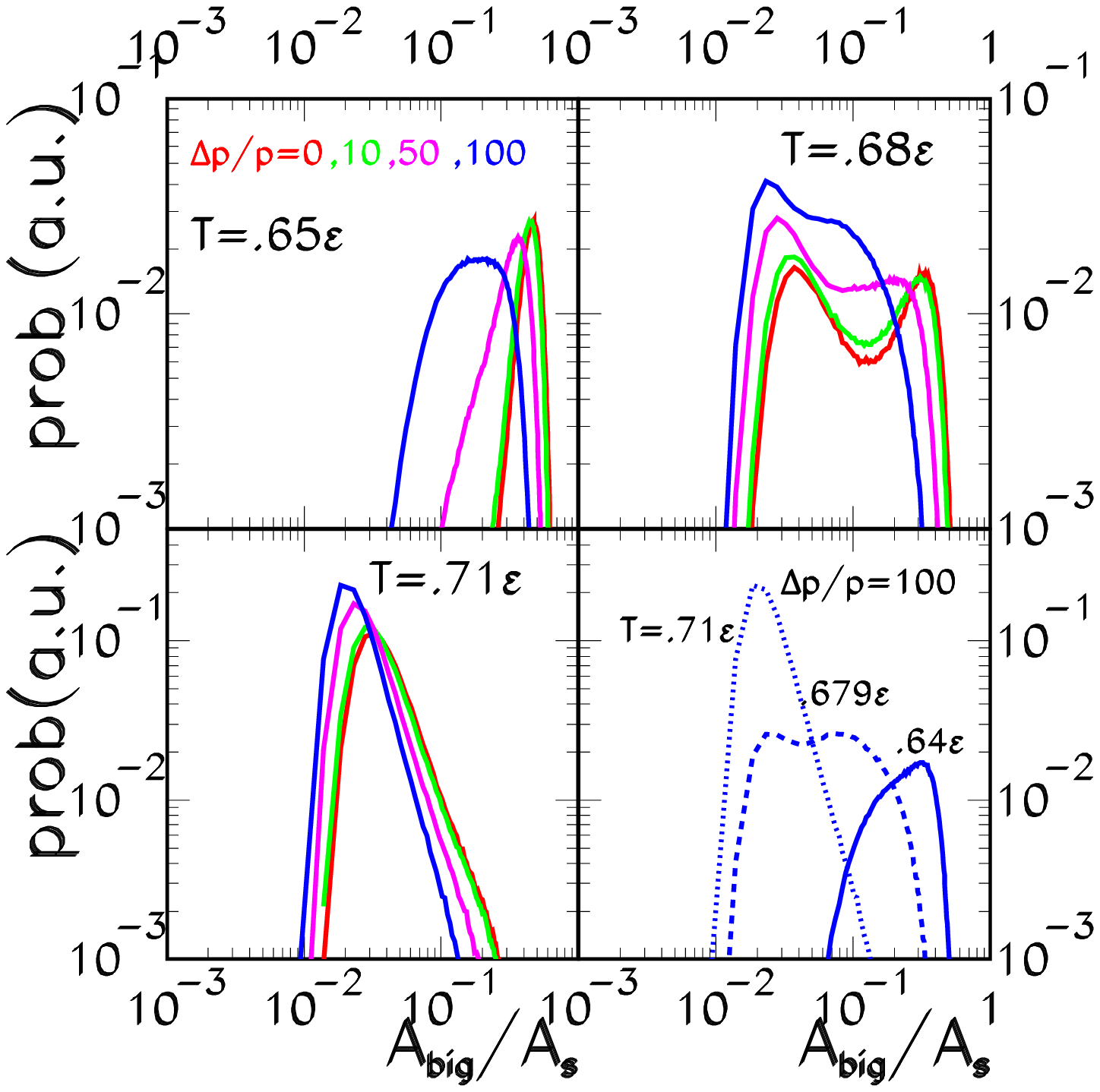}
\end{center}
\vspace{-0cm}
\caption{Similar to figure \ref{fig:gulmi} but an aligned momentum component has been added for 
each particle. The added aligned momentum is 0, 10, 50 or 100 percent of the thermalized one.
The last insert corresponds to the 100 percent case at three different temperatures.
Such calculations which include an out-of equilibrium component are presented in reference\cite{Chomaz}.}
\label{fig:gulmi2}
\end{minipage}
\end{figure}

An example is given in figure \ref{fig:gulmi} obtained in the lattice-gas approach 
(see for instance ref\cite{Gulmi}\cite{Chomaz}).
For temperatures outside of a small range around the transition temperature, the size of the largest fragment
exhibits single-humped distributions whereas two peaks of the same height are obtained at the transition temperature
value. This result which is obtained for an equilibrated system in the canonical isobar ensemble remains approximately 
valid even if part of the available energy is not thermalized. This is shown in figure 
\ref{fig:gulmi2} 
in which an aligned momentum has been randomly added for each particle, which 
simulates the dynamical effects corresponding to the memory of the aligned beam
momentum in the entrance channel. Several percentages of aligned momenta have been considered ranging from zero percent (purely
thermal situation) to 100 percent (same amount of aligned and thermalized momentum). The bimodality signal is 
robust in the sense that it is observed even if a sizeable amount of the total available momentum is 
still aligned along the beam direction, with only a slight change of the deduced transition temperature.
Only if the aligned momentum is as important as the thermal one (lower right in the figure) the distribution 
is weakly bimodal and mimics a flat distribution expected for a continuous transition\cite{Binder,Botet}.

Guided by these considerations, preliminary evidences of bimodal behaviours in some static observables 
have been  obtained in previous data concerning 
the systems Ni+Au\cite{Bell}, Ni+Ni\cite{Laut}, and Xe+Sn\cite{Bord3}. In none of these cases, 
a clear bimodality was observed in the distribution of the heaviest fragment charge. 
A possible reason is that references \cite{Bell,Laut,Bord3} concern central collisions. In such cases, 
the event selection is closer to a microcanonical sorting, 
with a selection of a narrow range of excitation energy: if the transition is characterized by a finite latent heat,
the two peaks that can be interpreted as precursors of the two coexisting phases\cite{Gulmi} should be 
associated to two different energies. The signal which would naturally appear if the system was in contact with a heat 
bath is expected to be quenched if the experimental sorting constrains too strongly the deposited excitation energy.

In order to mimic a canonical sorting in the data selection, one can select
semi-peripheral collisions which are mostly binary. 
In these collisions, it is possible to isolate a quasi-projectile (QP) and a quasi-target (QT). 
The fluctuations in the sharing of the dissipated energy between these two partners 
may be sufficient to explore the two phases.  
In the present work, we focus on such peripheral and semi-peripheral collisions. 

\section{The experiment}

The data have been obtained by the Indra-Aladin collaboration (Indra
at GSI). The Indra set up is described in references\cite{Pouthas1,Pouthas2}, while details of 
the measurements performed at the GSI laboratory and of the analysis and
calibration procedures may be found in \cite{Lefevre,Lukasik,Trz,Turzo}.
 The studied systems are Xe+Sn and Au+Au from 60 to 100
MeV/u. The events were registered if at least four modules fired. 
In this section the question of the event sorting is first addressed. 
Then we will check to which extent it is possible to isolate properly the QP 
from the QT and any mid-rapidity emission.

\subsection{Event sorting}\label{sorting}

In order to separate the QP and QT contributions, 
the momentum tensor calculated from $Z\geq 3$ products
has been diagonalized event by event. The main axis of the tensor is in most cases 
close to the beam axis. 
The particles and fragments attributed to the QT (resp. QP) are 
those which are backward (resp. forward) emitted along the main tensor axis. 
Event sorting has been achieved from the transverse energy $Etrans$ of light
charged particles (LCP) on the QT side calculated in this frame. A selected $Etrans$ value 
corresponds to a selected violence of the collision.

\begin{figure}[htb]
\begin{center}
\includegraphics[width=11cm]{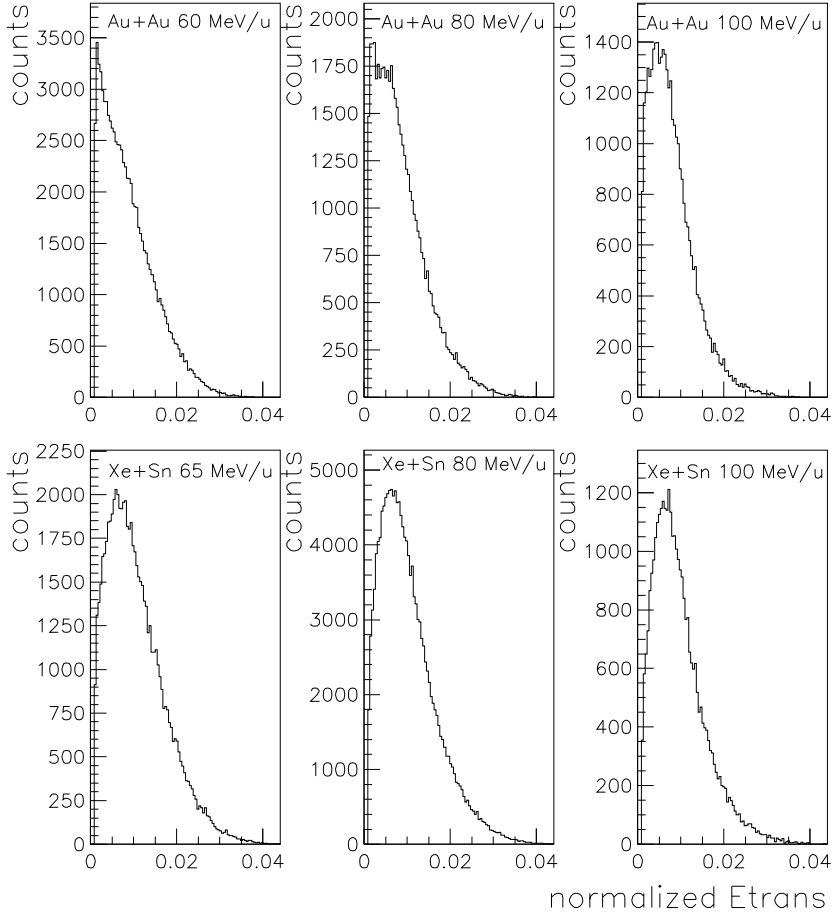}
\caption{$Etrans$ distributions for the two studied systems at the three
indicated energies. For the abscissa $Etrans$ has been divided 
by the total mass number of the system (projectile + target), and by the incident 
energy in MeV/u.}
\label{fig:etrans_dist}
\end{center}
\end{figure}

\begin{figure}[htb]
\begin{center}
\includegraphics[width=9cm]{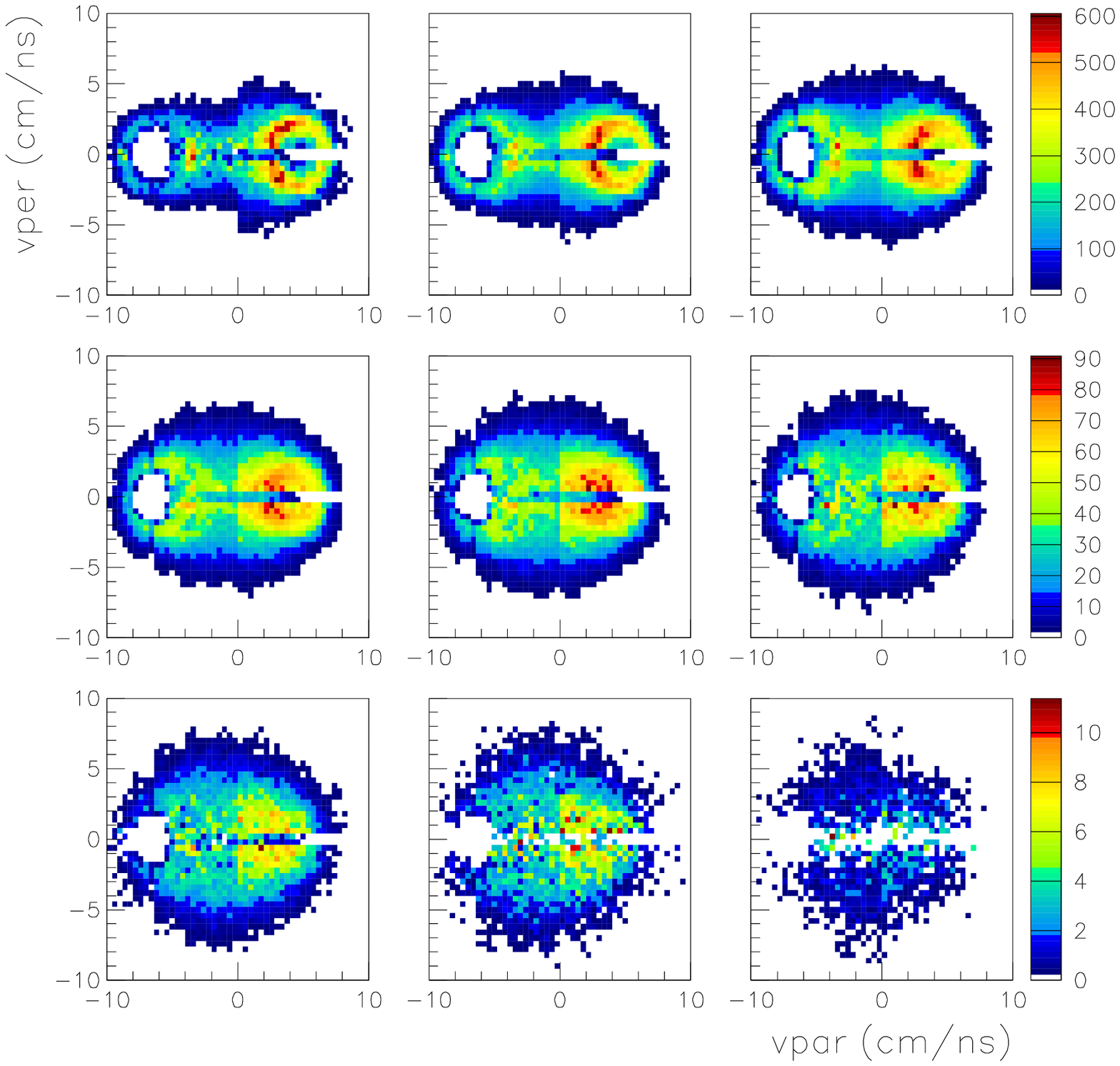}
\caption{Alpha particle centre-of-mass $v_{par}-v_{per}$ plots for nine $Etrans$ bins covering the 
whole $Etrans$ distribution of figure \ref{fig:etrans_dist}. Xe+Sn system at
80 MeV/u.The width of each $Etrans$ bin is 100 MeV i.e. 0.4 MeV/u if $Etrans$ is normalized on the total system 
mass as in figures \ref{fig:sautAu} and \ref{fig:sautXe}} 
\label{fig:vpar_vper_alpha}
\end{center}
\end{figure}

The $Etrans$ sorting has the advantage
of minimizing autocorrelations between the sorting variable and
observables, since the sorting particles are only loosely correlated to the
particles and fragments considered to get the QP properties. 
From the experimental point of view, 
the detection efficiency is good for LCP in the whole space since the LCP detection thresholds 
are small. 
Hence, the LCP transverse energy on the QT side was well measured. 
On the other hand, the detection efficiency on the QP side (velocities exceeding the c.m. 
velocity) is good for both the LCP
and the fragments since all the laboratory velocities are far above 
the detection thresholds for any product. 
In order to ensure an almost complete QP reconstruction, 
any event for which the forward detected charge was below
80 percent of the projectile charge has been rejected. This selection rejected mainly 
peripheral events for which the QP residue was emitted in the forward detector opening. 

The $Etrans$ distributions, normalized 
to the size of the system and to the laboratory bombarding energy in MeV/u, for different systems are 
shown in figure \ref{fig:etrans_dist}. They are similar 
and they extend on the same abscissa range whatever the system and the incident energy. 
The differences between the systems are mainly due to the 
loss of efficiency for very peripheral collisions when the quasi-projectile residue escapes 
in the detector forward hole (these events are rejected because of the completeness detection 
criterion).
This efficiency loss is stronger for the lighter system because QP fission has a negligible probability 
in the Xe mass range:
the QP loss in the forward hole is indeed reduced when fission occurs because the fission fragment 
emission angles are
generally pushed away from very forward directions.

The similarity of the various normalized-$Etrans$ distributions indicates that the energy damping (or the stopping power) 
does not depend strongly on the mass number nor on the available energy in this bombarding energy region. 
It is in agreement with the conclusions of reference\cite{Esca}.  
The largest values reached by the normalized $Etrans$ are about 0.04.
They can be compared with the limiting value of $1/12 \sim 0.08$ which would correspond (for such symmetrical systems) 
to a complete fusion followed by vaporization (with only LCPs as decay products).

\subsection{Kinematical separation between QP and QT decay products} 
This paper is devoted to the QP decay study. Hence, it must be verified
to which extent the QP and QT decay products can really be separated. 
This question can be answered looking at figure \ref{fig:vpar_vper_alpha} in which typical 
$v_{par}-v_{perp}$ 
invariant cross section plots are shown for alpha particles emitted in 80 MeV/u Xe on Sn collisions. 
$v_{par}$ and $v_{per}$ are the parallel and perpendicular velocity components in the reaction centre of mass frame.
Similar distributions are obtained for the other systems considered in this paper.
The events have been sorted in nine $Etrans$ 
bins. The distributions clearly exhibit two sources 
for peripheral and semi-peripheral collisions at least up to the fourth bin. For the highest $Etrans$ values
the distinction between a QP and a QT source becomes somewhat artificial.

A careful observation of figure \ref{fig:vpar_vper_alpha} indicates that alpha particles are not isotropically emitted 
from the QP and QT. Instead more alphas are emitted at mid-rapidity in the neck region. 
This result is a general feature in intermediate energy nucleus-nucleus collisions\cite{Dempsey,Lefort,Chimera}
and one may worry about the influence of a mid-rapidity emission on the results discussed in this paper. 
This will be discussed in section \ref{mid-rapid-influence}.

\begin{figure}[htb]
\begin{minipage}[t]{65mm}
\begin{center}
\includegraphics[width=6.5cm]{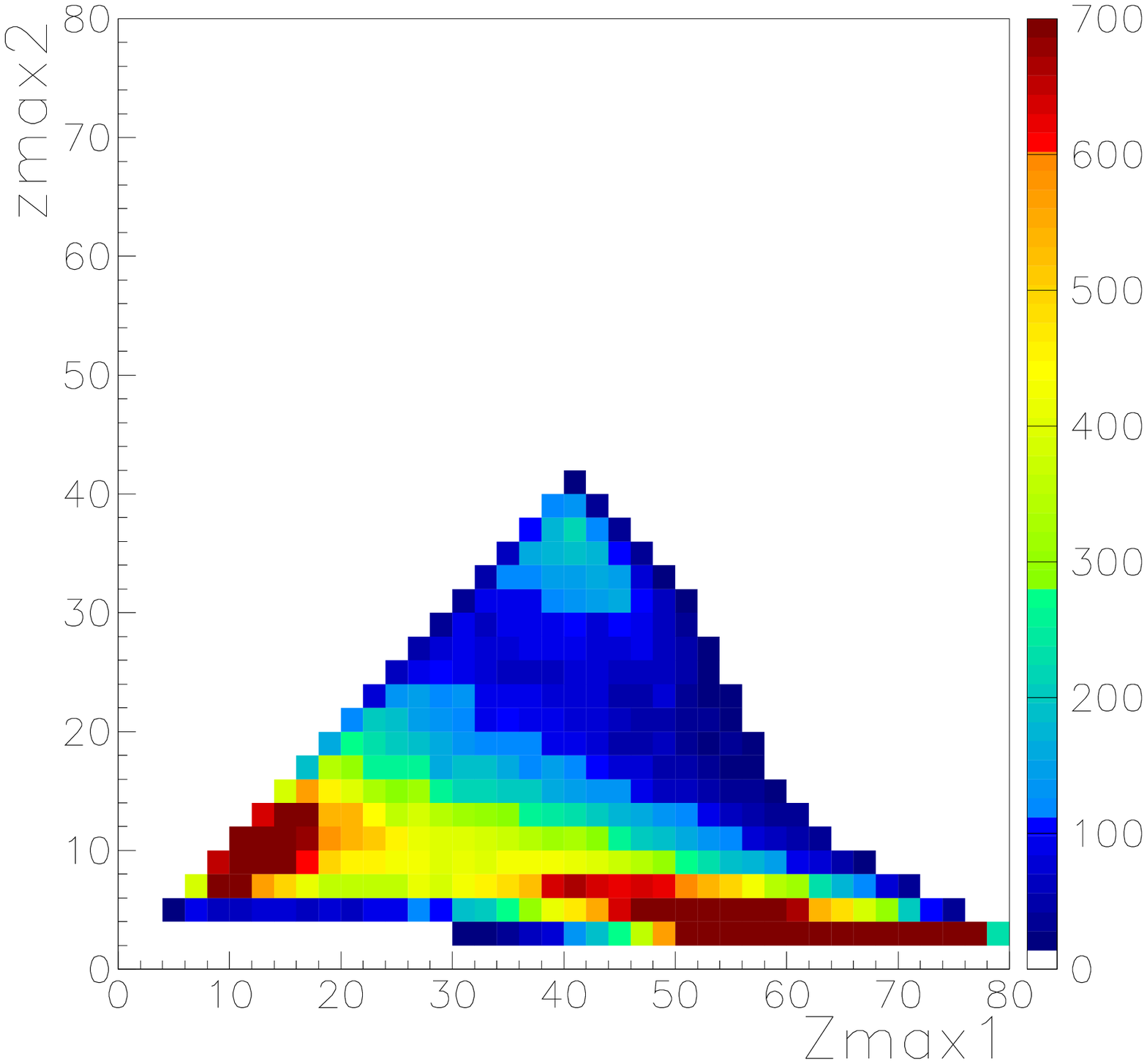}
\end{center}
\vspace{-0cm}
\caption{Correlation between the second largest charge $Zmax_2$ and the largest
charge $Zmax_1$ for the Au+Au system at 80 MeV/u.}

\label{fig:zmax-zmax-1-Au}

\end{minipage}
\hspace{\fill}
\begin{minipage}[t]{65mm}
\begin{center}
\includegraphics[width=6.5cm]{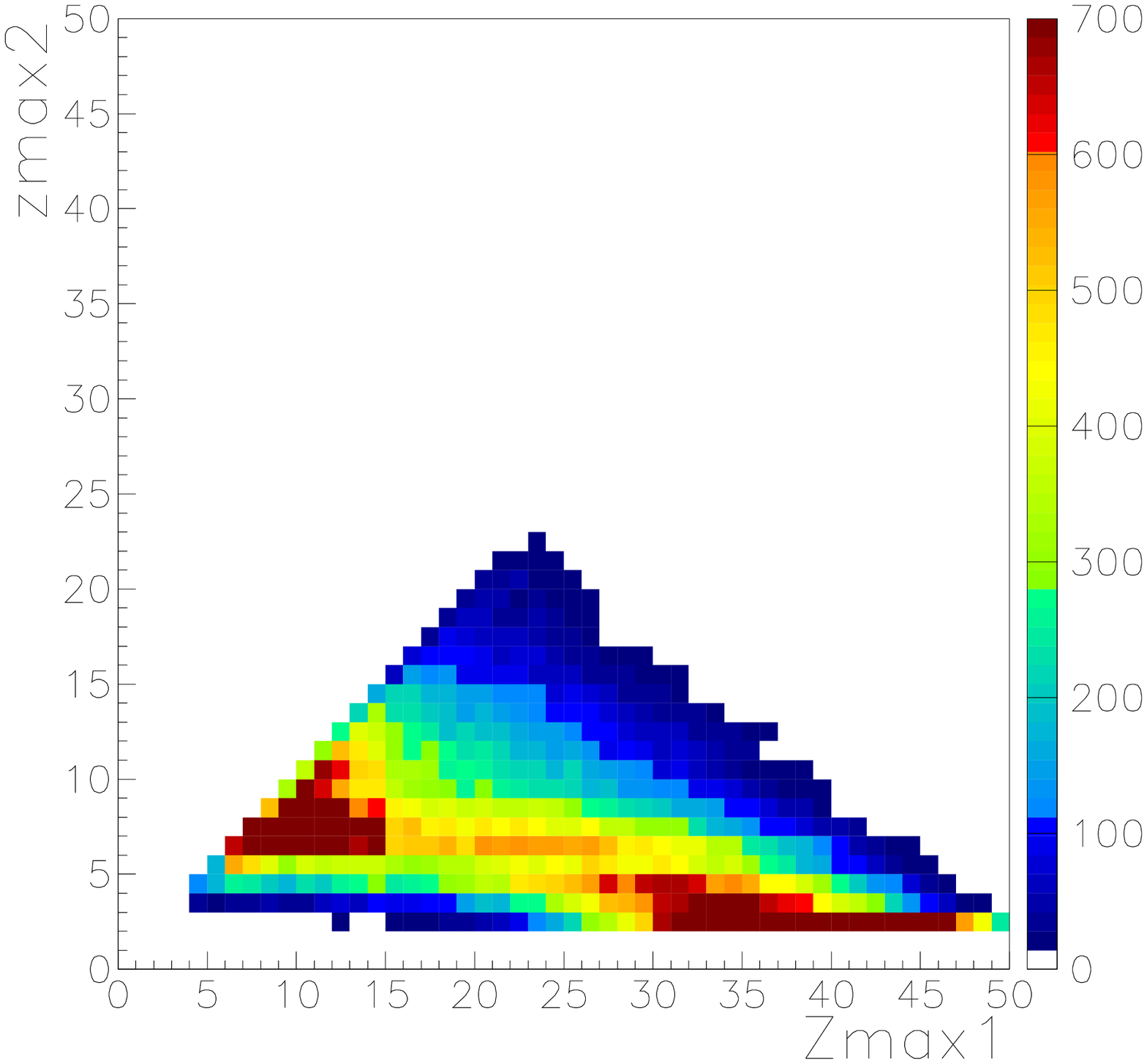}
\end{center}
\vspace{-0cm}
\caption{Correlation between the second largest charge $Zmax_2$ and the largest
charge $Zmax_1$ for the Xe+Sn system at 80 MeV/u.}
\label{fig:zmax-zmax-1-Xe}
\end{minipage}
\end{figure}

\section{QP observables}
Following the discussion of the introduction, a natural choice of the order parameter is the size (charge) of the 
largest fragment $Zmax_1$. 
The asymmetry $Z_{asym}$ between the two heaviest products $Zmax_1$ and $Zmax_2$
(see below equation \ref{zasym} for its precise definition) will also be studied.

In figures \ref{fig:zmax-zmax-1-Au} and  
\ref{fig:zmax-zmax-1-Xe}, the correlations
between  $Zmax_1$ and $Zmax_2$ are shown for the Au+Au and Xe+Sn systems, respectively, at 80 MeV/u. 
All the retained (complete) events are included in the picture. For the Au+Au system, 
the binary fission of the quasi-gold nucleus can be recognized for $Zmax_1$ and $Zmax_2$ values 
around 40 charge units. This fission contribution is negligible for the Xe case. For both reactions
one recognizes also two dominant classes of events. Residue like events correspond to a large $Zmax_1$ and a small 
$Zmax_2$. Multifragmentation events correspond to small values for both $Zmax_1$ and $Zmax_2$.
An important feature is that there is not a continuous evolution from one class of events to the other.
Instead, they are widely separated showing that they correspond to two clearly distinct 
mechanisms. 

For Au QP, the presence of the fission channel perturbs the evolution of $Z_{asym}$ since this variable 
exhibits similar values for multifragmentation and 
binary fission. In order to avoid this difficulty, we applied a specific treatement to the fission events 
observed 
for the Au case. In effect, they belong also to the normal compound nucleus decay event class and can 
be classified together with the residue events. For 
this purpose, we defined $Z_{max}$ as the sum of the two fission fragment charges when fission was recognized. 
Fission events have been selected from the charge product of the two fission fragments: 
charge product exceeding 900\cite{Dagos} (see also figure \ref{fig:zmax-zmax-1-Au}). For these events, $Z_{max}$ is the sum of the two heaviest products and 
we defined $Z_{sec}$ as the third heaviest product charge. For the other Au events 
and for any event in the Xe case (for which the fission contribution is negligible),
one has $Z_{max}$ = $Zmax_1$ and $Z_{sec}$ = $Zmax_2$ . Then the asymmetry parameter is 
defined by:
\begin{equation}
Z_{asym} = \frac {Z_{max}-Z_{sec}}{Z_{max}+Z_{sec}}
\label{zasym}
\end{equation}
In order to avoid spurious peaks due to integer number effects we have added a random number between
 -0.5 and +0.5 to $Z_{max}$ and $Z_{sec}$. Thus $Z_{asym}$ can exceed slightly the range [0-1].

\begin{figure}[htb]
\begin{minipage}[t]{65mm}
\begin{center}
\includegraphics[width=6.5cm]{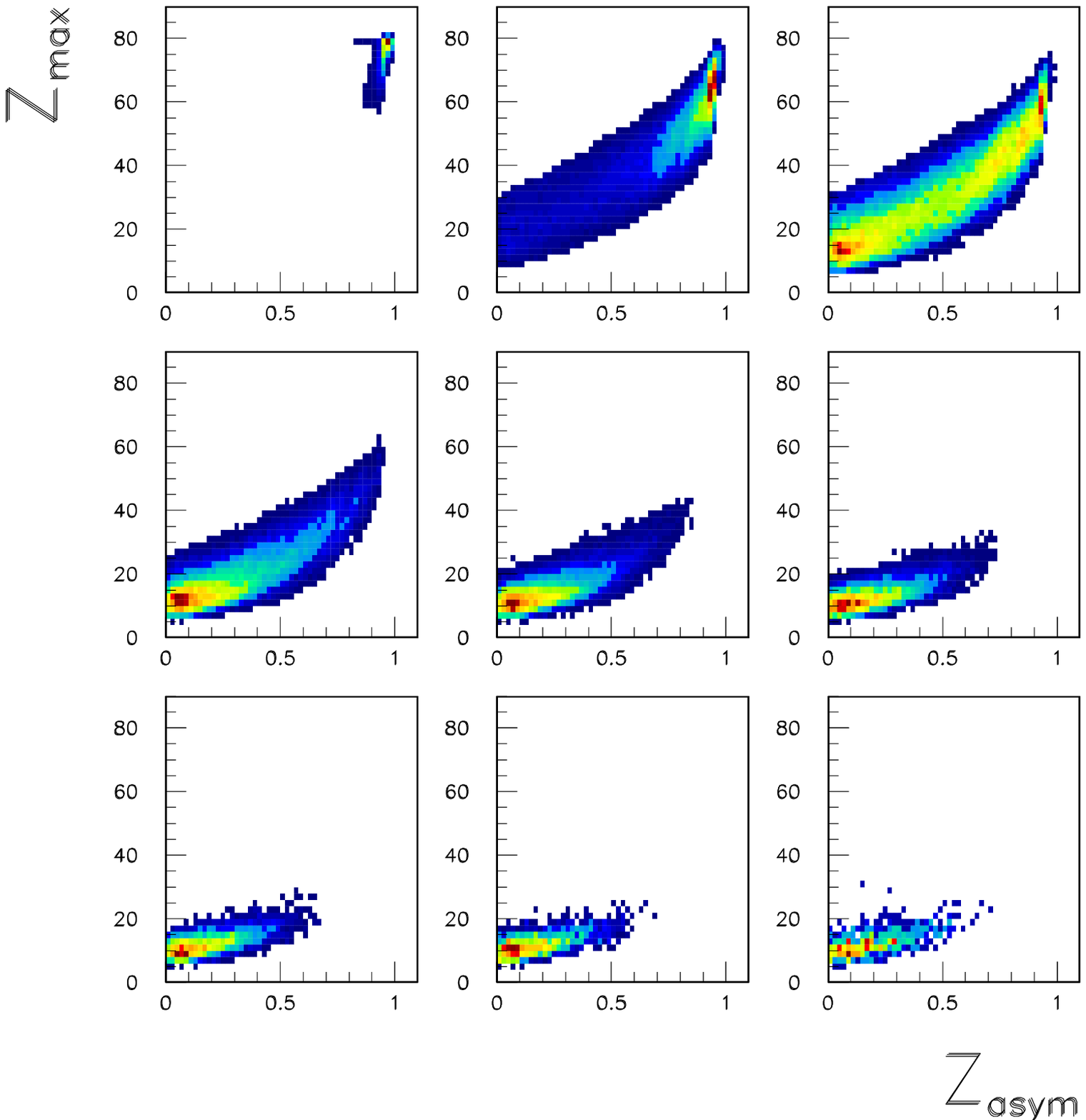}
\end{center}
\vspace{-0cm}
\caption{$Z_{max} -Z_{asym}$ plots for the nine $Etrans$ bins. Au+Au system at
  80 MeV/u.}

\label{fig:Zasym-zmax-Au}

\end{minipage}
\hspace{\fill}
\begin{minipage}[t]{65mm}
\begin{center}
\includegraphics[width=6.5cm]{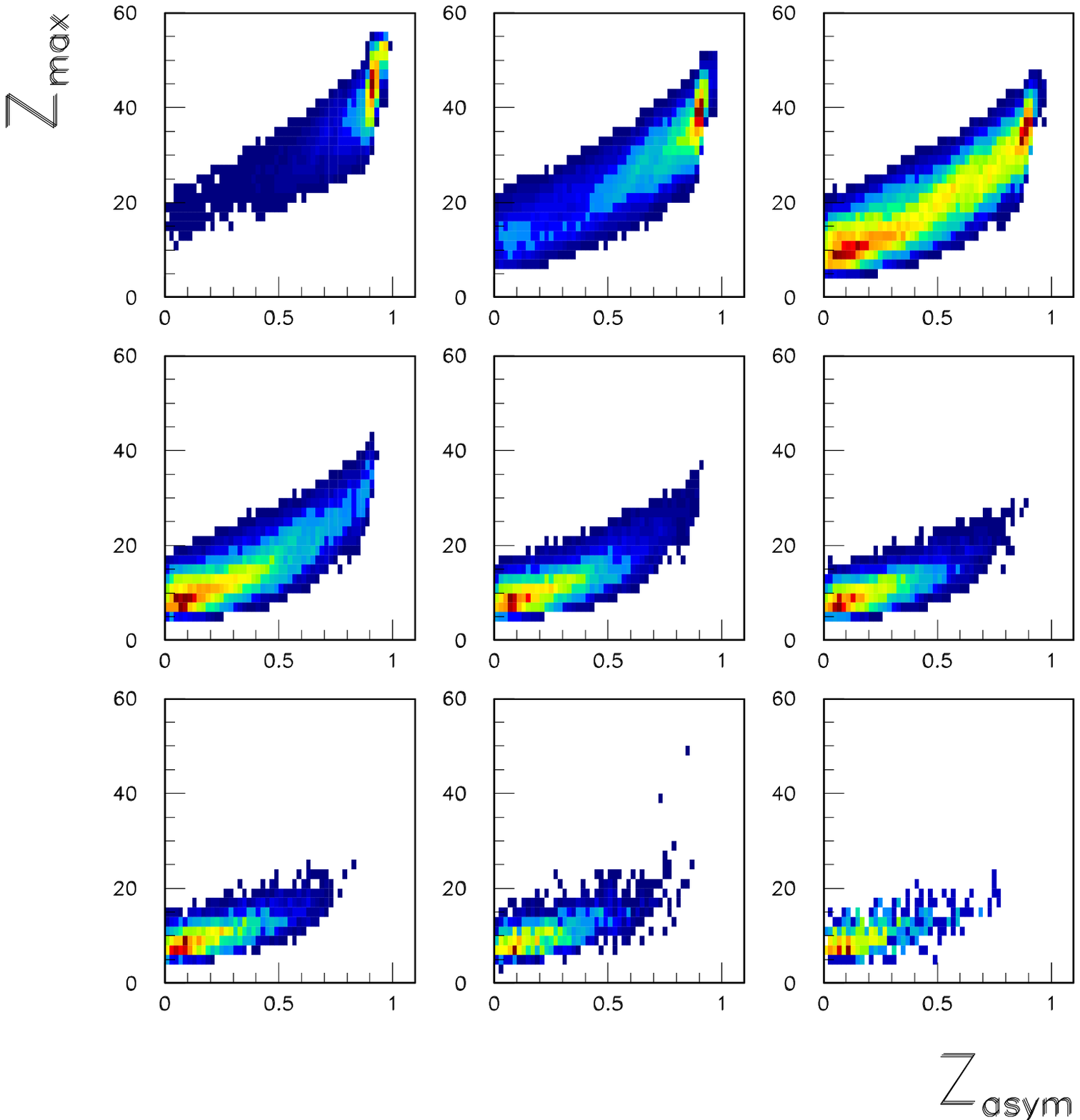}
\end{center}
\vspace{-0cm}
\caption{$Z_{max} -Z_{asym}$ plots for the nine $Etrans$ bins. Xe+Sn system at
  80 MeV/u.}
\label{fig:Zasym-zmax-Xe}
\end{minipage}
\end{figure}

\begin{figure}[htb]
\begin{minipage}[t]{65mm}
\begin{center}
\includegraphics[width=6.5cm]{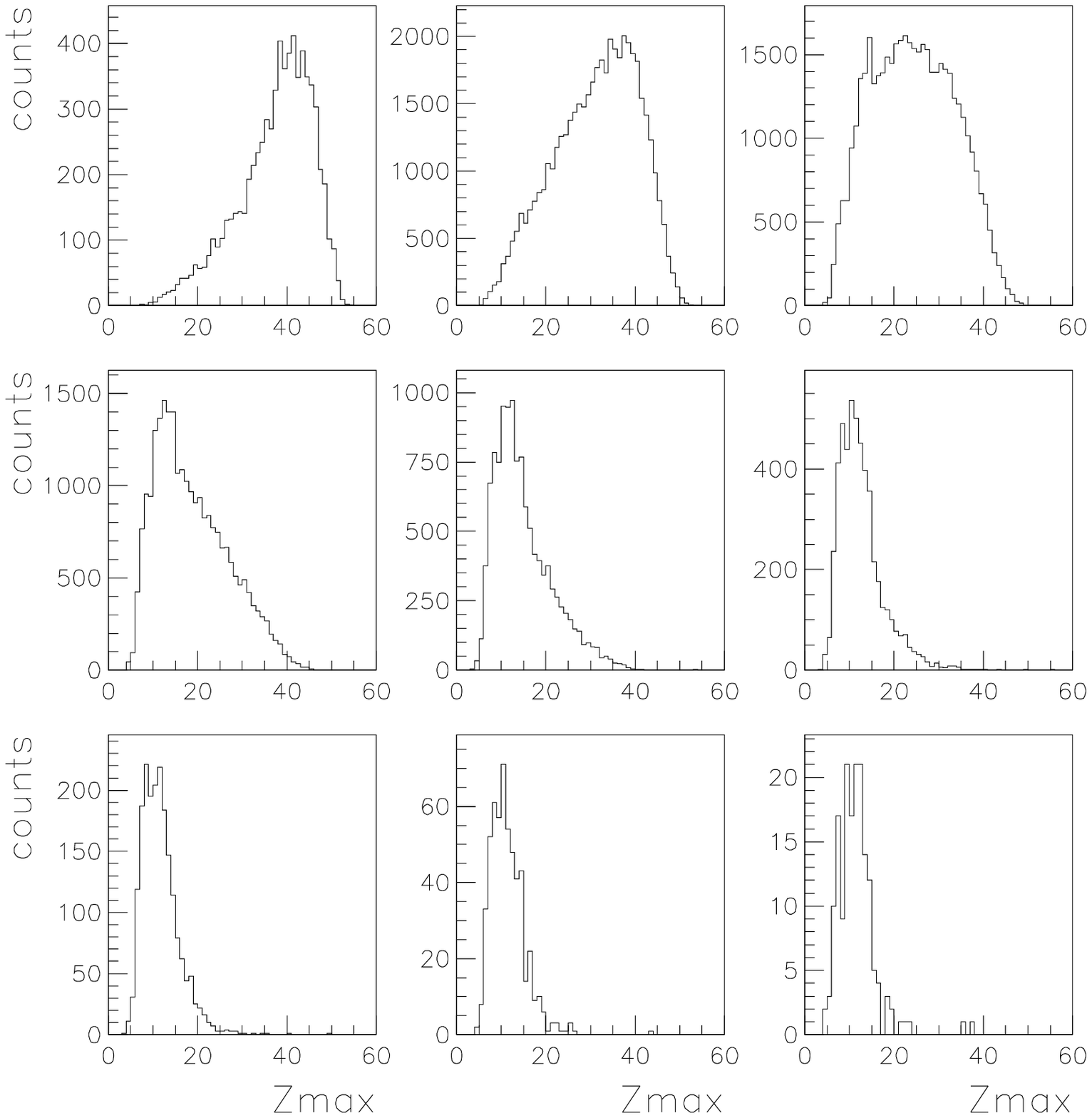}
\end{center}
\vspace{-0cm}
\caption{$Zmax$ distributions for the 9 $Etrans$ bins. Xe+Sn system at
  80 MeV/u.}

\label{fig:Zmaxproj-Xe}

\end{minipage}
\hspace{\fill}
\begin{minipage}[t]{65mm}
\begin{center}
\includegraphics[width=6.5cm]{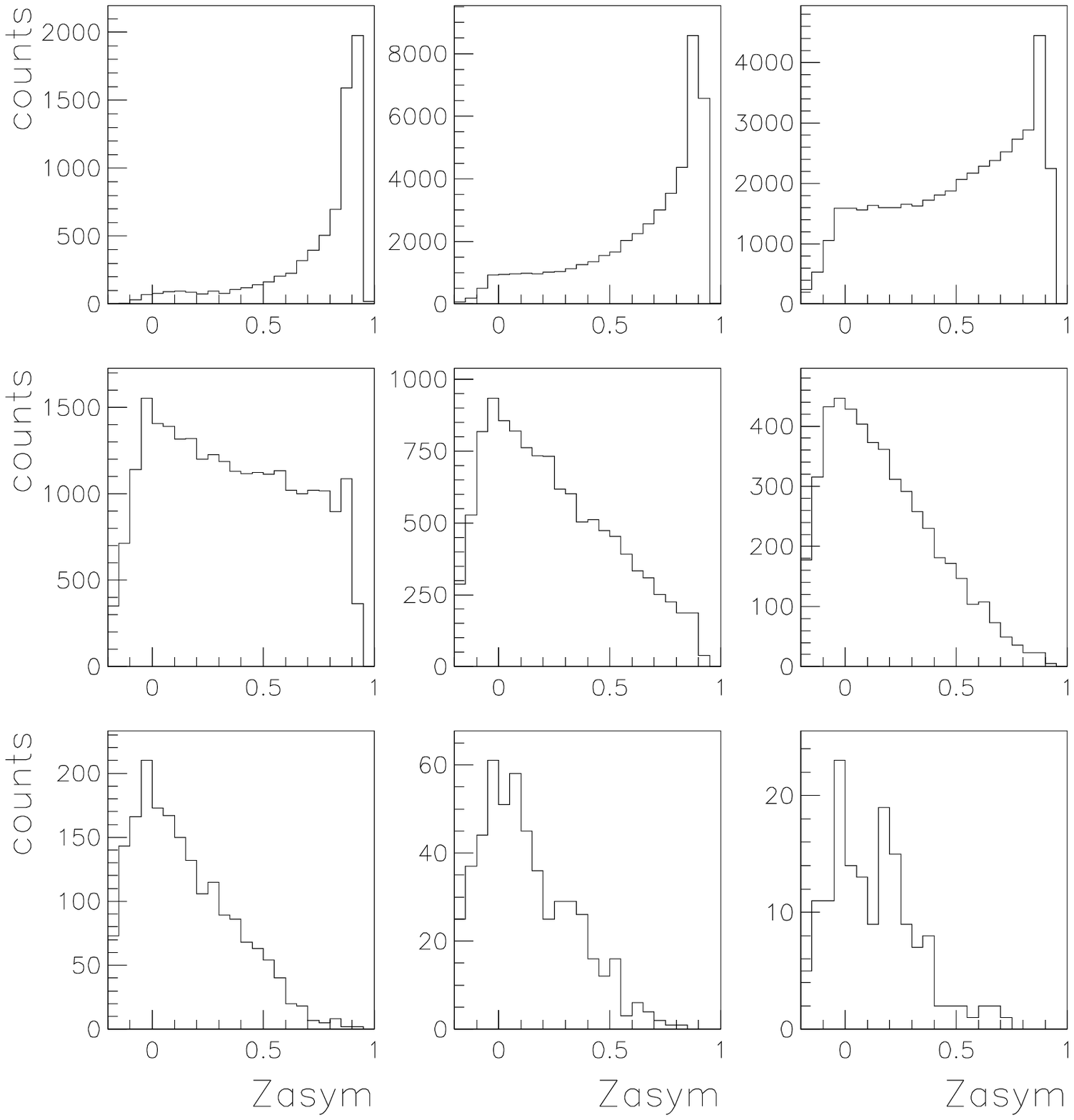}
\end{center}
\vspace{-0cm}
\caption{$Z_{asym}$ distributions for the 9 $Etrans$ bins. Xe+Sn system at
  80 MeV/u.}
\label{fig:Zasymproj-Xe}
\end{minipage}
\end{figure}

\section{Bimodality observation}\label{observ}
Let us consider now the evolution of the correlation between $Z_{max}$ and $Z_{asym}$ 
when the events are sorted according to the transverse energy $Etrans$ defined in section \ref{sorting}.
Figures \ref{fig:Zasym-zmax-Au} and \ref{fig:Zasym-zmax-Xe} correspond to the systems Au+Au and Xe+Sn, respectively, 
at 80 MeV/u. They show the correlations between 
the asymmetry variable $ Z_{asym} $ and the heaviest product charge $Z_{max}$ for the nine $Etrans$ bins already
used in figure \ref{fig:vpar_vper_alpha}. 
Two behaviours are clearly evidenced in the figures. In the first bins, 
the heaviest decay product is generally a residue (large $Z_{max}$ and $Z_{asym}$ value close to unity). 
By contrast, the most violent (central) collisions correspond to small values of $Z_{max}$ and $Z_{asym}$ 
associated with multi-fragment emission (such observations are quite in agreement with the results of ref \cite{Kreutz}).
In between, in the third bin, both behaviours are observed: the distribution is bimodal. Similar results have been 
obtained by Ma et al\cite{Ma} on a lighter system. Here we have such an observation on medium and large mass systems. 
With the definition of $Z_{max}$ as the sum of the two fission fragment charges (see the previous section),
the fission event behaviour is quite similar to the residue one. We have checked that, if one does not reconstruct 
the parent nucleus charge before fission, these events correspond to an additional bump in figure 
\ref{fig:Zasym-zmax-Au} widely separated from the residue and the fragmentation bumps. It does not modify the 
bimodality behaviour observed in the third bin. 

\begin{figure}[htb]
\begin{minipage}[t]{65mm}
\begin{center}
\includegraphics[width=6.5cm]{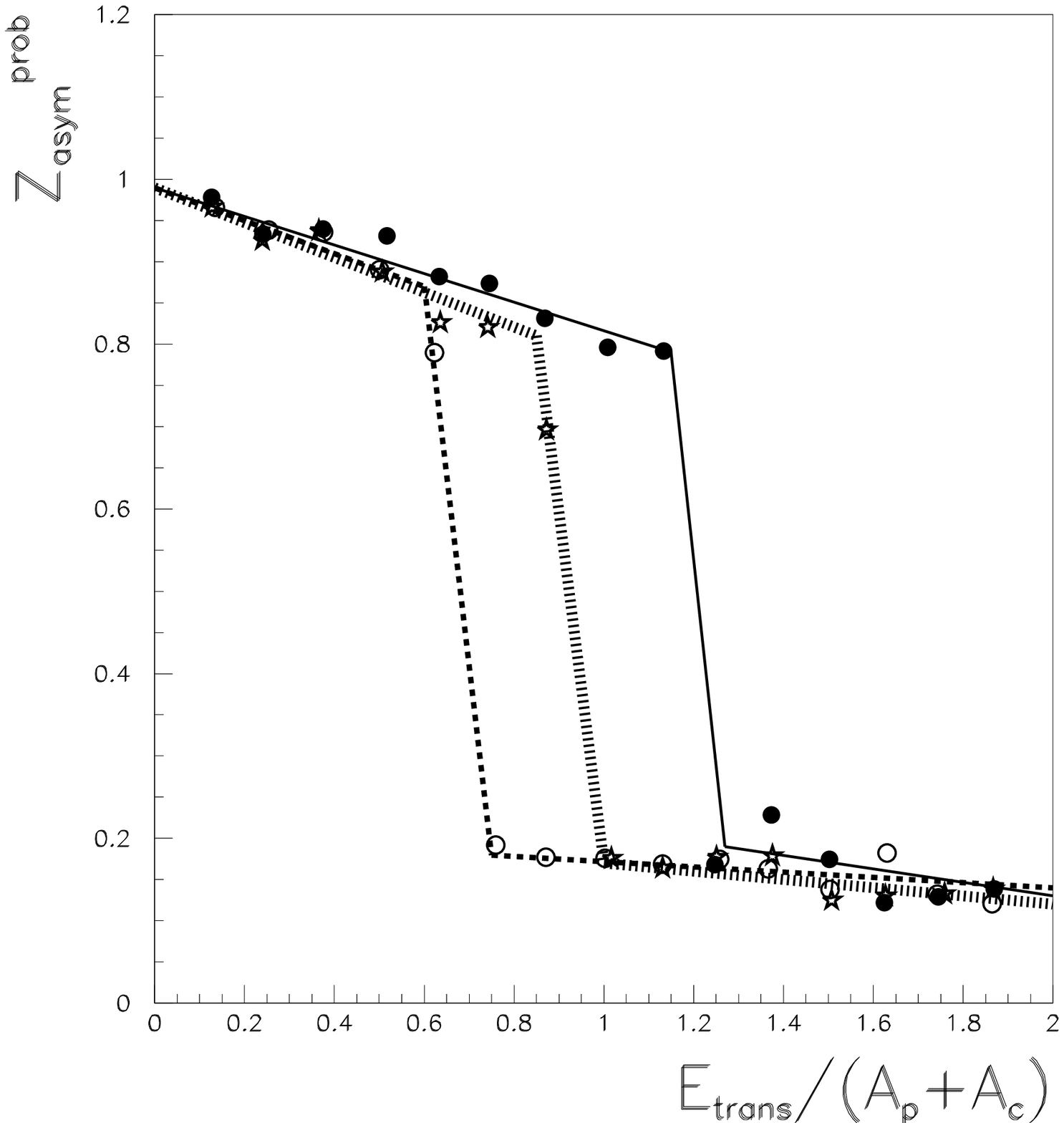}
\end{center}
\vspace{-0cm}
\caption{System Au + Au at 60, 80, 100 MeV/u (respectively open points, stars and black points).
  Evolution with $Etrans$ of the most probable value of the $Z_{asym}$ distributions.
  In this figure, $Etrans$ has been divided by the total mass number of the system 
  (projectile + target). It has not been divided by the incident 
  energy as in figure \ref{fig:etrans_dist}.}

\label{fig:sautAu}

\end{minipage}
\hspace{\fill}
\begin{minipage}[t]{65mm}
\begin{center}
\includegraphics[width=6.5cm]{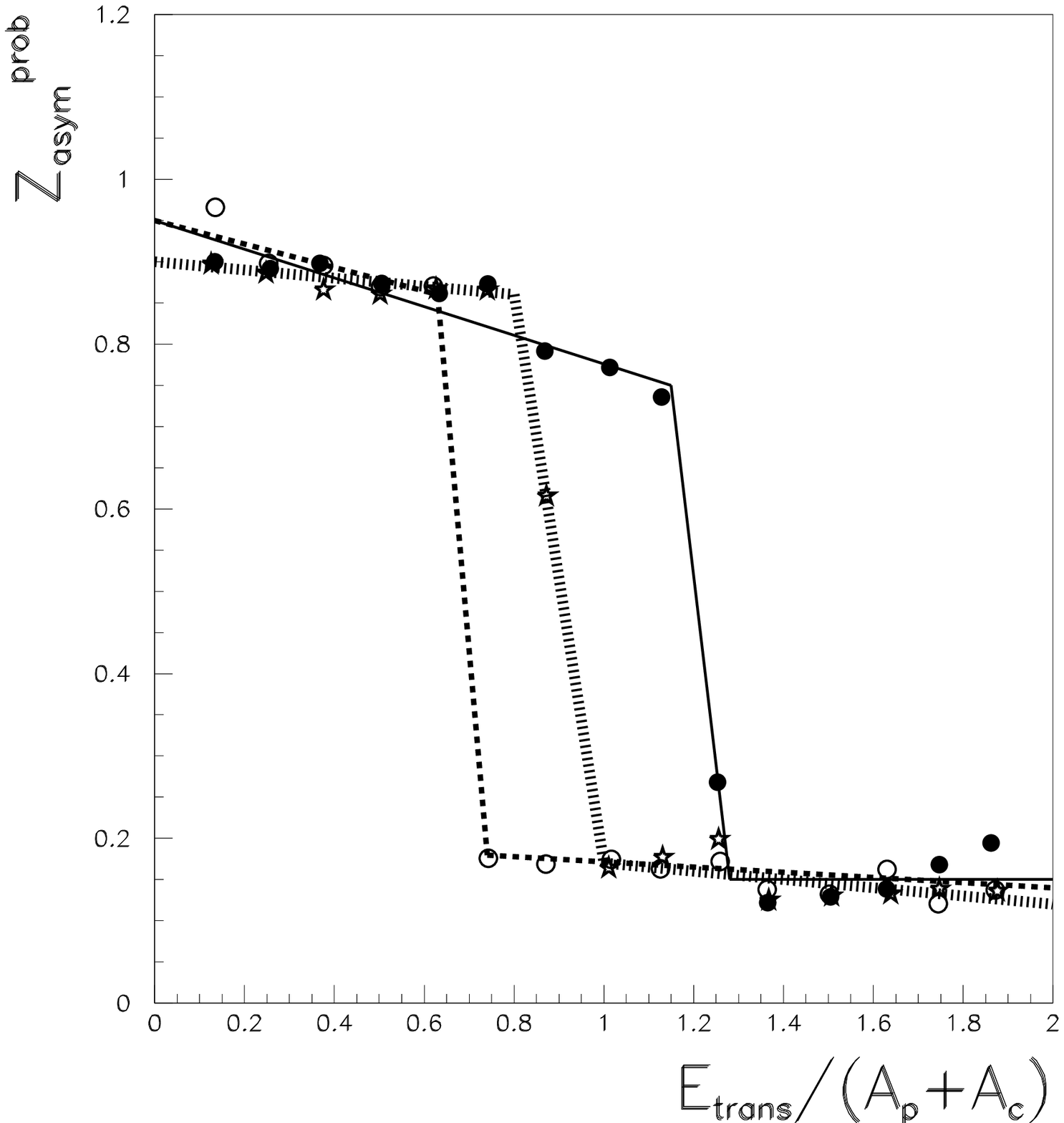}
\end{center}
\vspace{-0cm}
\caption{Similar to figure \ref{fig:sautAu} but for the systems Xe + Sn at 65, 80 and 100 MeV/u.
  The abscissa corresponding to the transitions are similar for Au + Au and Xe + Sn system. 
  This scaling indicates that the transverse energy per nucleon is a key parameter.}
\label{fig:sautXe}
\end{minipage}
\end{figure}

Although the event probability distribution in the third bin exhibits a minimum in the representation of 
figures \ref{fig:Zasym-zmax-Au} 
and \ref{fig:Zasym-zmax-Xe}, this minimum (convex shape) is not apparent on the projections on the two axes, 
as shown in figures \ref{fig:Zmaxproj-Xe} and \ref{fig:Zasymproj-Xe} for the Xe+Sn system at 80 MeV/u.  
Nevertheless, these figures indicate that the most probable values of the $Z_{max}$ or $Z_{asym}$ 
distributions change dramatically between the third and the fourth  bins. This evolution is summarized 
in figures \ref{fig:sautAu} and \ref{fig:sautXe} for the two studied systems and the three bombarding energies. 
In these figures, the abscissa is the transverse energy sorting parameter $Etrans$ divided by the system size 
(sum of the projectile and target masses). 

The abrupt change of the order parameter together with the wide and almost structureless distribution at the 
transition point shown in figures \ref{fig:Zasym-zmax-Au}-\ref{fig:Zasymproj-Xe} are rather suggestive of a continuous 
(second order) transition point. However, as we have discussed in section 1, such a behaviour is also consistent 
with a first order transition polluted by strong out of equilibrium effects, 
or important deformations due to sorting constraints\cite{Gulmi2005}: 
if the sorting variable is too strongly connected to the order parameter, its
bimodal character is naturally suppressed.

\begin{figure}[htb]
\begin{center}
\includegraphics[width=9cm]{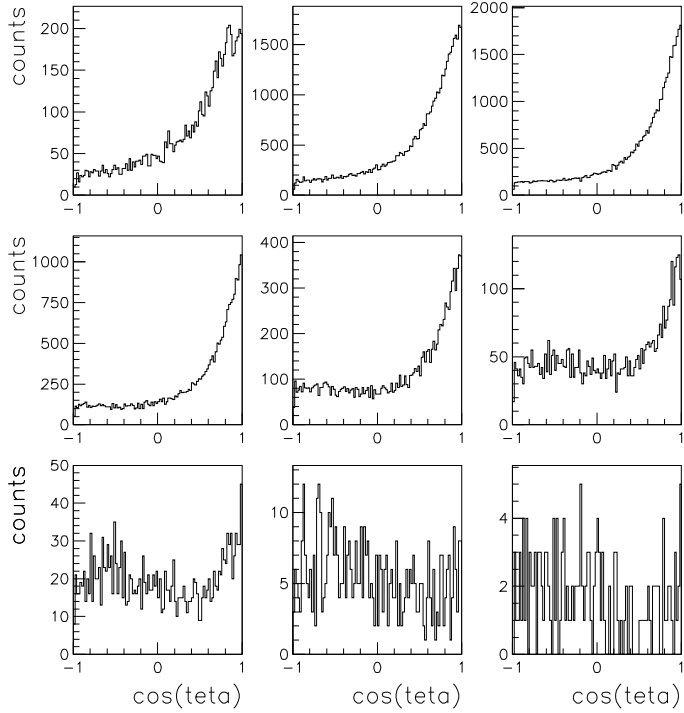}
\caption{Angular distributions of the heaviest QP fragment in the QP frame. The abscissa 
is the cosine of the angle between the QP direction in the centre of mass system and
the $Zmax1$ fragment direction in the QP frame. The events have been sorted according to the 
$Etrans$ variable. Most of the heaviest fragments are forward emitted. Xe+Sn system at
80 MeV/u.} 
\label{fig:disangzmax}
\end{center}
\end{figure}

\section{Is bimodality due to dynamics?}\label{mid-rapid-influence}

In comparing figures \ref{fig:sautAu} and \ref{fig:sautXe}, one may
conclude that the transverse energy per nucleon at which the transition occurs
does not depend on the system size. A similar kind of universality has also been pointed out in reference 
\cite{Schu}. Such a result
is expected if the transition reflects the energy density involved in the reaction. 
The transition energy 
for a definite system evolves when the bombarding energy increases: it
is roughly proportional to the incident energy. This result is not expected 
if $Etrans$ reflects the thermal energy stored in the quasi-projectile.

However, $Etrans$ reflects both the dissipated energy and 
preequilibrium effects. When one increases the bombarding energy, the preequilibium contribution
increases thus inducing a larger value of $Etrans$ to reach the same thermal energy. 
A similar behaviour has been 
evidenced from the calculations discussed in figure \ref{fig:gulmi2}. 
To test such an interpretation, we have to introduce an additional selection of the events in order 
to decrease the influence of preequilibrium effects. Section \ref{dynamic} is devoted to the corresponding data analysis.  

One may also consider that the  
transition (where bimodality is observed) corresponds to an intermediate impact parameter 
which would be about the same whatever the bombarding energy is. If for this centrality a massive 
fragment emission from the neck region sets in, the sudden change of the $Z_{max}$ distribution could be
related to a sudden decrease of the QP source mass, i.e. to a geometrical effect. 
This interpretation is discussed in sections \ref{mid-rapid-substr} and \ref{mid-rapid-model}.
.

\subsection{The role of midrapidity emission.}\label{mid-rapid-substr}

Dynamical effects were clearly recognized in nucleus-nucleus collisions in the intermediate energy regime. 
We already noticed in figure \ref{fig:vpar_vper_alpha} the possible role of neck emission which has been
extensively studied in reference\cite{Lukasik,Colin,Plagnol}. A generalization of the results of reference\cite{Colin} is 
shown in figure \ref{fig:disangzmax} where the angular distributions of the heaviest fragment ($Zmax1$) are drawn 
for various $Etrans$ bins. The abscissa is the cosine of its emission angle in the QP frame. For this 
analysis, the QP centre of mass has been calculated including all products with $Z \geq 2$ emitted forward in
the ellipsoid frame in order to 
be able to treat events with a residue and no intermediate mass fragment (IMF: $Z \geq 3$). The distributions 
would be flat for isotropic emission which is not the case. Mostly, the heaviest fragment 
is forward emitted in the QP frame. This feature is a clear entrance channel memory: 
full equilibrium has not been achieved at least for most events and light IMF or LCP are emitted at mid-rapidity
between the QP and QT. What is the role of this mid-rapidity emission? Does it contribute significantly to the 
sudden increase of the fragment multiplicity when passing from the residue-like class to the multifragmentation 
one?

\begin{figure}[htb]
\begin{center}
\includegraphics[width=9cm]{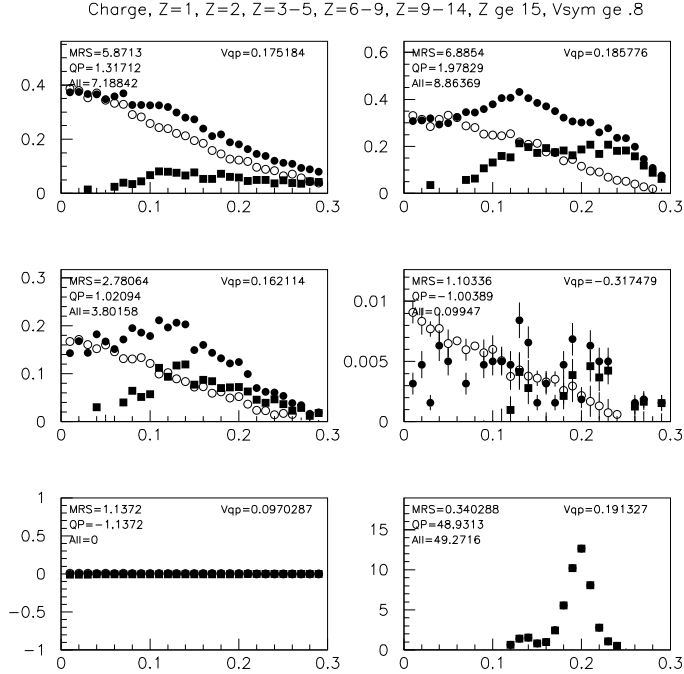}
\caption{Charge density distributions for various IMF charges as a function of the parallel velocity to the beam in the c.m. frame 
in c units. The full and open circles are the distributions for $Etrans$ bins 3 and 9 respectively. They have 
been normalized for zero velocity. The black squares correspond to the difference between the two
previous distributions. For the third bin the events with $Z_{asym}$ larger than 0.8 have been selected. 
System Au + Au at 100 MeV/u.}
\label{fig:mirap1}
\end{center}
\end{figure}

In order to evaluate the role of mid-rapidity we performed an analysis based
on the hypothesis that mid-rapidity emission comes mainly from the overlap region between the 
projectile and the target and that its kinematical properties are the same for all the impact parameters.
This assumption is in agreement with the basic hypothesis of the HIPSE model\cite{Durand} which is discussed in 
section \ref{mid-rapid-model}.

Central collisions are those for which the overlap between the projectile and the target is the largest: 
the overlap or participant zone is in this case the unique source for particle emission. The corresponding velocity properties have 
been assumed to be the reference properties of mid-rapidity emission at any impact parameter
(similar parallel velocity distribution shapes for any impact parameter). They have been summarized 
in plotting the charge density $\rho $ as a function of the velocity parallel to the beam $v_{par}$ in the centre of 
mass frame.

\begin{figure}[htb]
\begin{center}
\includegraphics[width=9cm]{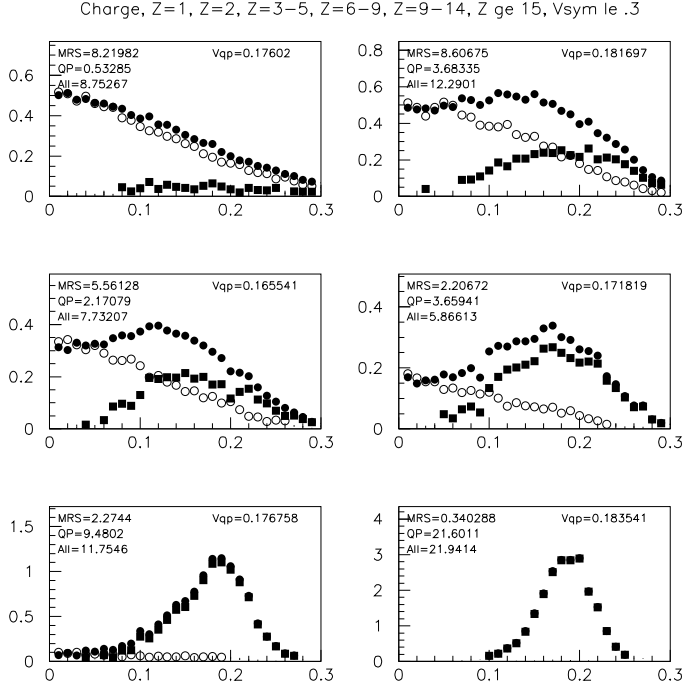}
\caption{Similar to figure \ref{fig:mirap1} but the events selected for the third bin are multifragmentation events
($Z_{asym}$ lower than 0.2).}
\label{fig:mirap2}
\end{center}
\end{figure}

The Au + Au system at 100 MeV/u will be taken as an example. Central collisions have been associated
with the largest $Etrans$ bin. As a matter of fact, one may see in figure \ref{fig:vpar_vper_alpha} that the 
$vpar-vper$ plots for the ninth bin is dominated by mid-rapidity emission 
with a negligible contribution of QP or QT emissions. 
Now, we focused on the third $Etrans$ bin where the bimodality signal is seen. In figures \ref{fig:mirap1} (resp. 
\ref{fig:mirap2}) we have plotted the $\rho (vpar)$ distributions for several IMF charges and for $Z_{asym}$ 
larger than 0.8 for residue-like selection (resp. lower than 0.2 for multifragmentation). The black dots correspond 
to the data. The open ones are the corresponding 
reference distribution obtained for central events (ninth bin). The two distributions have been normalized at zero 
velocity which means that we assumed that the contribution of the QP at mid-rapidity was negligible. The difference 
between the two distributions (black squares) is attributed to the sequential QP decay. In figure \ref{fig:mirap1} 
the statistics is very poor for $Z$ between 6 and 14 because very few IMF are emitted when a residue is released. 
On the contrary, a larger number of IMF is obtained when one selects multifragmentation (figure \ref{fig:mirap2}). 
The resulting QP contributions (black squares) for all charges are symmetric around the QP velocity derived from 
the bottom right frame as it is expected for 
late sequential decay. This means that any QP decay contribution with an influence from the QT 
partner, which would show a higher yield of particles in the backward direction\cite{Desouza,Yennello}, 
has been included 
in the mid-rapidity contribution. The contribution attributed to the QP in this analysis is hence underestimated.

\begin{table}[!h]
 \begin{center}
  \begin{tabular}{|c|c|c|c|c|c|c|c|}
   \hline
   \textit{$Z_{asym}$} & \textit{Z} & \textit{$M_{tot}$} & \textit{$M_{QP}$} & \textit{$M_{mi-rap}$} 
   & \textit{$Z_{tot}$} & \textit{$Z_{QP}$} & \textit{$Z_{mi-rap}$} \\
   \hline
   $> 0,8$ & LCP & 11.62 & 2.3 & 9.31 & 16.1 & 3.3 & 12.7  \\
   \hline
   $> 0,8$ & IMF & 2.14 & 1.36 & 0.77 & 53.2 & 47.8 & 5.3 \\
   \hline
   $> 0,8$ & Total & 13.76 & 3.66 & 10.08 & 69.2 & 51.1 & 18.1 \\
   \hline
   $< 0,2$ & LCP & 14.89 & 2.37 & 12.51 & 21.0 & 4.2 & 16.8  \\
   \hline
   $< 0,2$ & IMF & 5.06 & 3.00 & 2.06 & 47.3 & 36.9 & 10.4 \\
   \hline
   $< 0,2$ & Total & 19.95 & 5.37 & 14.57 & 68.3 & 41.1 & 27.2 \\
   \hline
  \end{tabular}
 \end{center}
 \caption{Repartition between the equilibrated QP and the mid-rapidity of the 
total available charge forward emitted in the centre of mass. The first three lines are for the residue-like events. 
The three last ones correspond to multifragmentation events. They all belong to the third $Etrans$  bin for which bimodality 
is observed. $M_{mi-rap}$, $M_{QP}$, $M_{tot}$ are the multiplicities respectively associated to mid-rapidity, QP emission, and 
any emission for LCP or IMF or any charged product. $Z_{mi-rap}$, $Z_{QP}$, $Z_{tot}$ are the corresponding charges.
Au + Au system at 100 MeV/u.} 
 \label{tab:tamain}
\end{table}

One may now deduce from the above treatment the minimum QP charge or the maximum mid-rapidity contribution in the
third $Etrans$ bin of interest, i.e. when bimodality is observed. The results are indicated in table 
\hbox{\ref{tab:tamain}}. The mid-rapidity contribution is never dominant and the mean 
IMF multiplicity attributed to the QP increases from 1.36 to 3 from residue like to multifragment events. This result
seems to indicate that the bimodality signal (disappearance of the residue and strong increase of IMF production)
is not dominated by mid-rapidity emission.

\subsection{Comparison with a dynamical model.}\label{mid-rapid-model}
\begin{figure}[htb]
\begin{center}
\includegraphics[width=9cm]{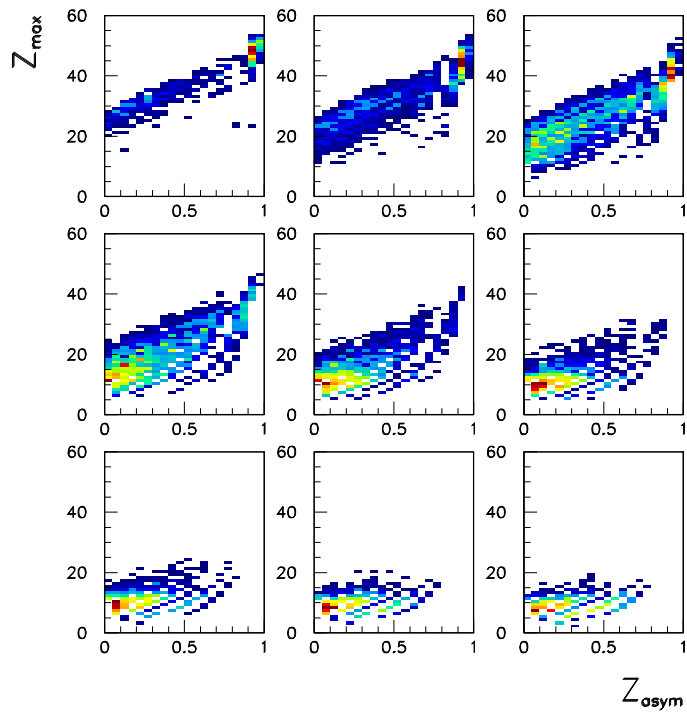}
\caption{$ Z_{asym} -Z_{max}$ plots for the 9 $Etrans$ bins as it is calculated by the HIPSE event generator. 
Xe+Sn system at 80 MeV/u.}
\label{fig:hipse1}
\end{center}
\end{figure}

\begin{figure}[htb]
\begin{center}
\includegraphics[width=9cm]{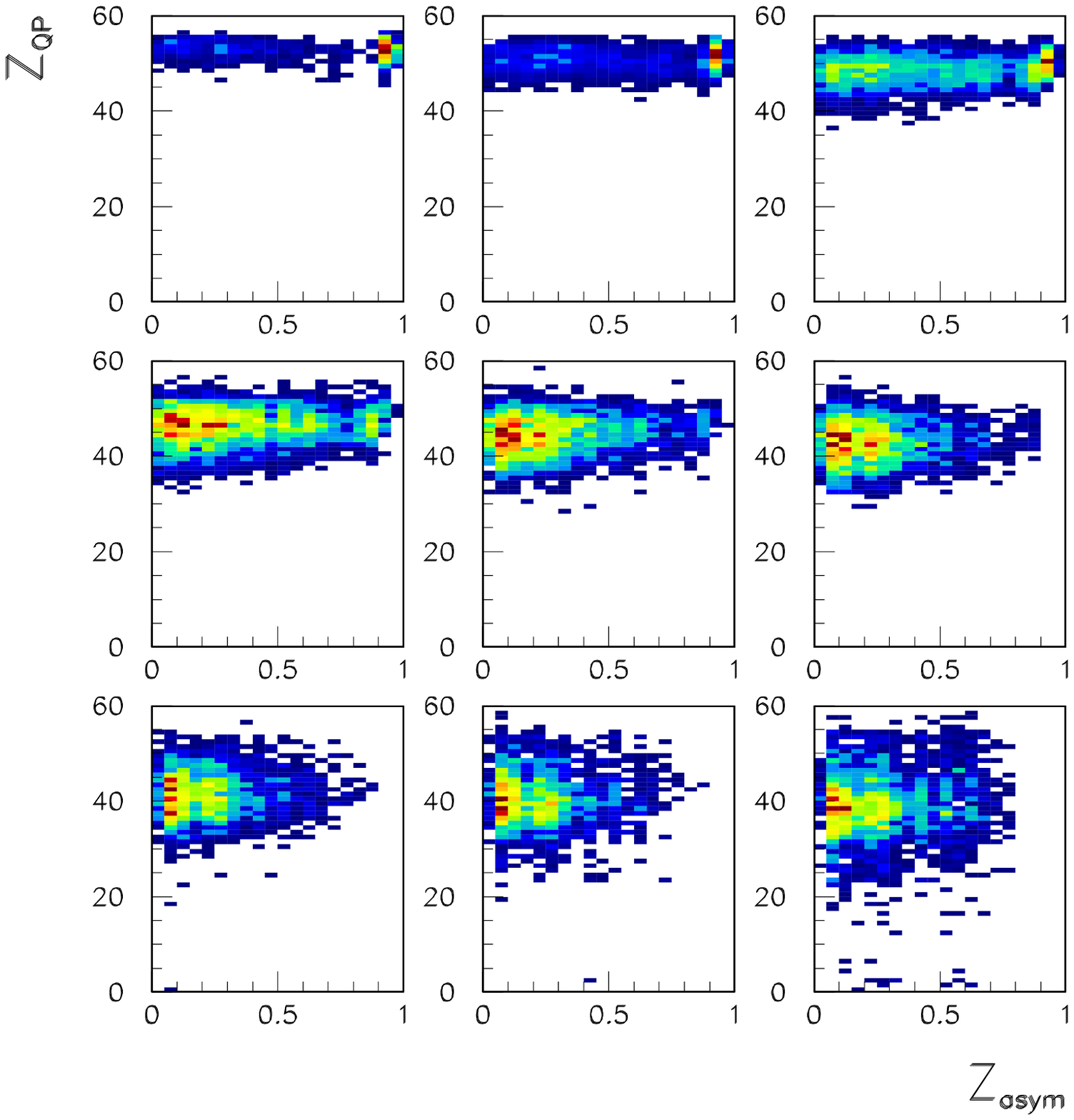}
\caption{$Z_{asym} -Z_{QP}$ plots for the 9 $Etrans$ bins as it is calculated by the HIPSE event generator. 
Xe+Sn system at 80 MeV/u.}
\label{fig:hipse2}
\end{center}
\end{figure}

The role of dynamics has been tested with the help of the event generator HIPSE\cite{Durand} in which the geometry 
of the collision is explicitly taken into account. For any impact parameter the overlap region leads to mid-rapidity 
emission, and contributes to QP excitation from nucleon-nucleon collisions. This is thus particularly suited to explore 
the puzzling property observed in
figures \ref{fig:sautAu} and \ref{fig:sautXe}: the bimodality transition is always observed for similar 
impact parameters corresponding to a $Etrans$  value which increases roughly proportionally with the beam energy. 
The parameters of the model have been adjusted in order to reproduce most of the experimental 
inclusive distributions (charge, angular and energy distributions) for Xe+Sn reactions between 25 and 50 MeV/u. 
The global experimental features are well reproduced by the model, as shown in figure \ref{fig:hipse1} 
for the Xe + Sn system at 80 MeV/u. This figure is similar to the experimental one (figure \ref{fig:Zasym-zmax-Xe})
and one recognizes bimodal distributions in the third and the fourth bins in very good agreement with the data. 
In the model, it is possible to go further and search the reason why the residue production vanishes when $Etrans$ 
is increased. 
The reason for this change is not directly the fact that a decrease of the impact parameter leads to a 
smaller QP spectator. This conclusion may be drawn from figure \ref{fig:hipse2} 
in which is shown the correlation between the initial hot QP size and $Z_{asym}$ for the nine $Etrans$ bins. It turns out that 
the initial QP size is always large. This is especially the case for bins 3 and 4, i.e. in the region where bimodality 
is observed in figure \ref{fig:hipse1}. This result means that it is the decay step which is responsible for the 
disappearance of the residue. It is found from the model that both the excitation energy and the angular momentum of the QP
become quite large
(3-5 MeV/u and 70$\hbar$) when the abrupt change in the decay mechanism is observed. 
In reference \cite{Lopez}, by using a sequential decay code, the authors conclude that the angular momentum is 
the key parameter. To progress on the understanding of the nature of the transition 
it will be necessary to perform similar 
simulations with a non-sequential decay code and, from the experimental point of view, to compare systems 
with very different values of angular momentum but similar excitation energies in 
the range of 3-5 MeV/u.

\begin{figure}[htb]
\begin{minipage}[t]{65mm}
\begin{center}
\includegraphics[width=6.5cm]{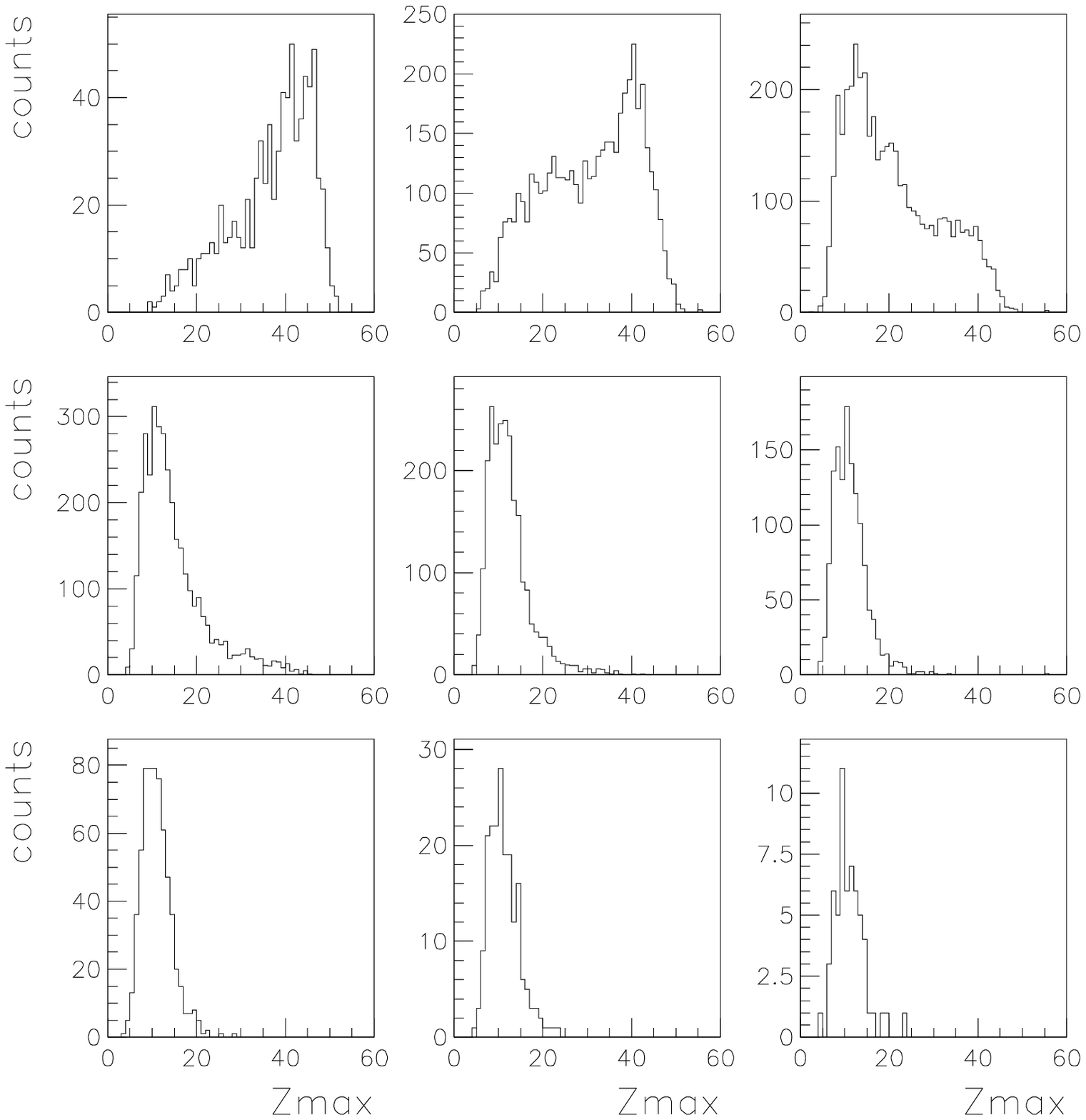}
\end{center}
\vspace{-0cm}
\caption{$Z_{max}$ distributions for the 9 $Etrans$ bins when the ''less dynamical events''
(see text) are selected. Xe+Sn system at
  80 MeV/u.}

\label{fig:Zmaxproj-Xe-equi}

\end{minipage}
\hspace{\fill}
\begin{minipage}[t]{65mm}
\begin{center}
\includegraphics[width=6.5cm]{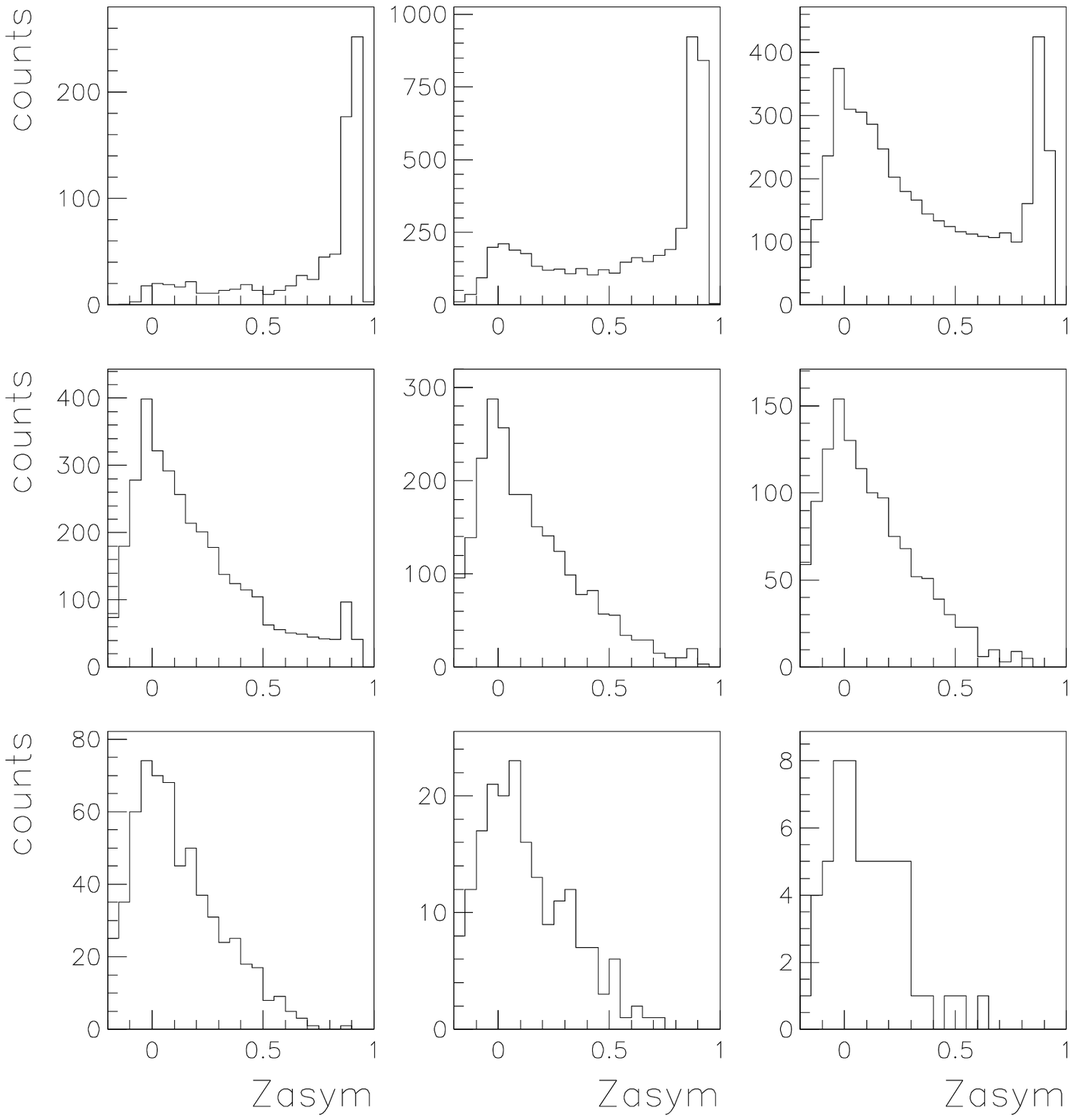}
\end{center}
\vspace{-0cm}
\caption{$Z_{asym}$ distributions for the 9 $Etrans$ bins. when the ''less dynamical events''
(see text) are selected. Xe+Sn system at 80 MeV/u.}
\label{fig:Zasymproj-Xe-equi}
\end{minipage}
\end{figure}

\subsection{Test of preequilibrium effects}\label{dynamic}
As shown in section \ref{mid-rapid-substr}, the angular distributions of the heaviest 
fragment, ($Zmax1$), are forward peaked (figure \ref{fig:disangzmax}), which
indicates a  memory of the entrance channel. 
Nevertheless, for some events, the heaviest fragment is  emitted backward in
the QP frame. The memory of the entrance channel is weaker for these events
which will thus be labelled ''less dynamical events''.
In figures \ref{fig:Zmaxproj-Xe-equi} and \ref{fig:Zasymproj-Xe-equi}, they have been isolated
in selecting $cos(\theta )$ values lower than $-0.4$ (see figure \ref{fig:disangzmax}). 
By comparison with those obtained when this condition 
is not required (figures \ref{fig:Zmaxproj-Xe} and \ref{fig:Zasymproj-Xe}),
two facts may be emphasized: 
firstly the bimodal behaviour is better evidenced 
when the ''less dynamical events'' are selected. 
This is especially the case for the $Z_{asym}$ variable 
which exhibits clearly a double humped distribution in the third $Etrans$ bin. 
On the other hand, the transition takes place for a lower $Etrans$ value for these selected events. 
All these features are expected if bimodality reflects a thermal behaviour. Indeed, lattice-gas  
calculations (figure \ref{fig:gulmi2}) indicate that the bimodality signal is attenuated 
by dynamical effects; the observed difference between the two samples of events may be explained in this framework. 
Moreover, when dynamical effects are present, 
we can expect the value of $Etrans$ at the transition point to be increased since the contribution of a
non-relaxed energy component does not modify significantly the bimodality signal.

\section{Transition excitation energies and temperature}\label{excit-ener}

\begin{figure}[htb]
\begin{minipage}[t]{65mm}
\begin{center}
\includegraphics[width=6.5cm]{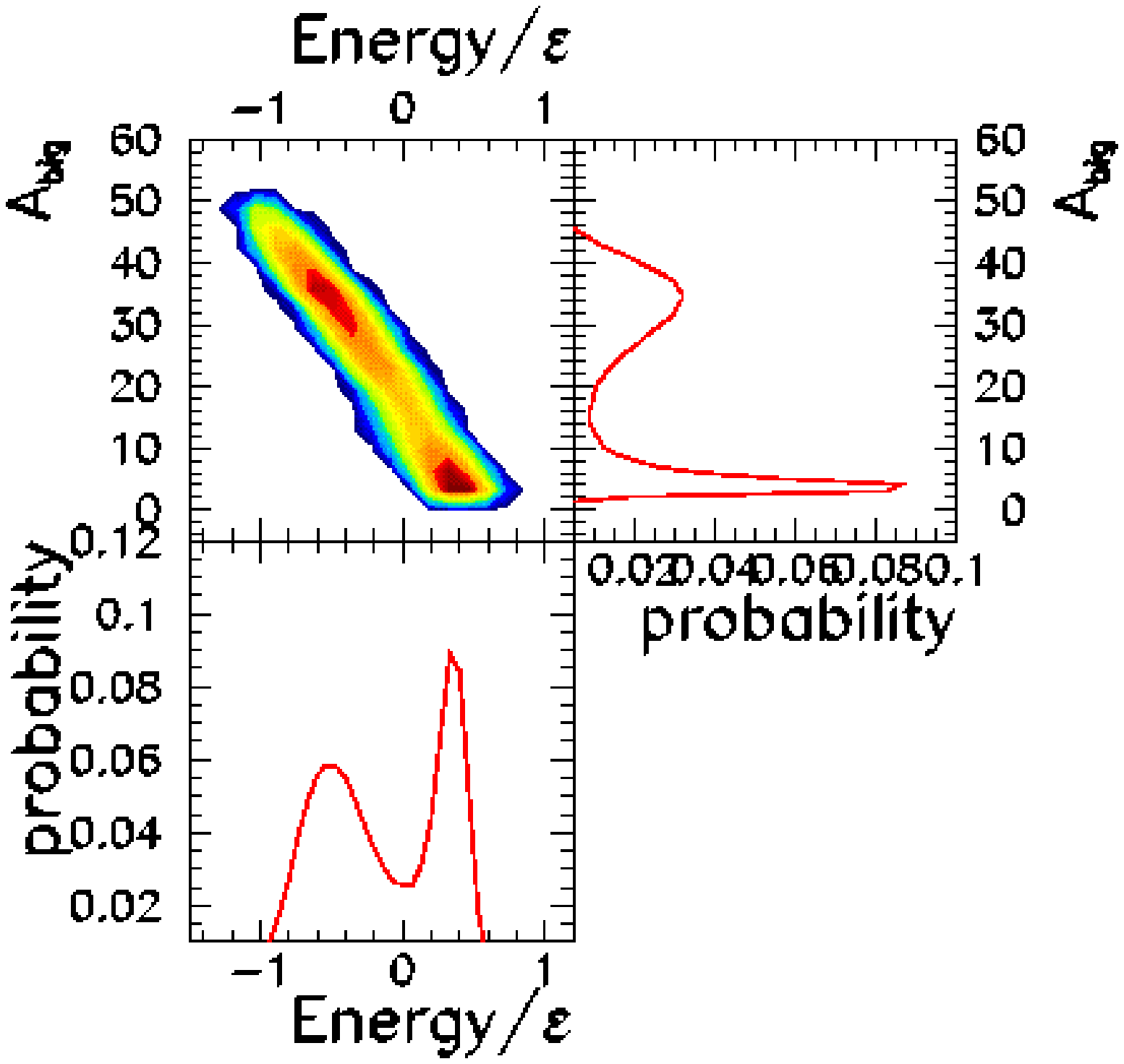}
\end{center}
\vspace{-0cm}
\caption{Correlation between the largest fragment size and the internal excitation energy 
from lattice-gas calculations close to the transition temperature (see ref \cite{Chomaz}). 
The projections are also shown. Bimodality is observed for both variables. }
\label{fig:gulmi_correl}
\end{minipage}
\hspace{\fill}
\begin{minipage}[t]{65mm}
\begin{center}
\includegraphics[width=6.5cm]{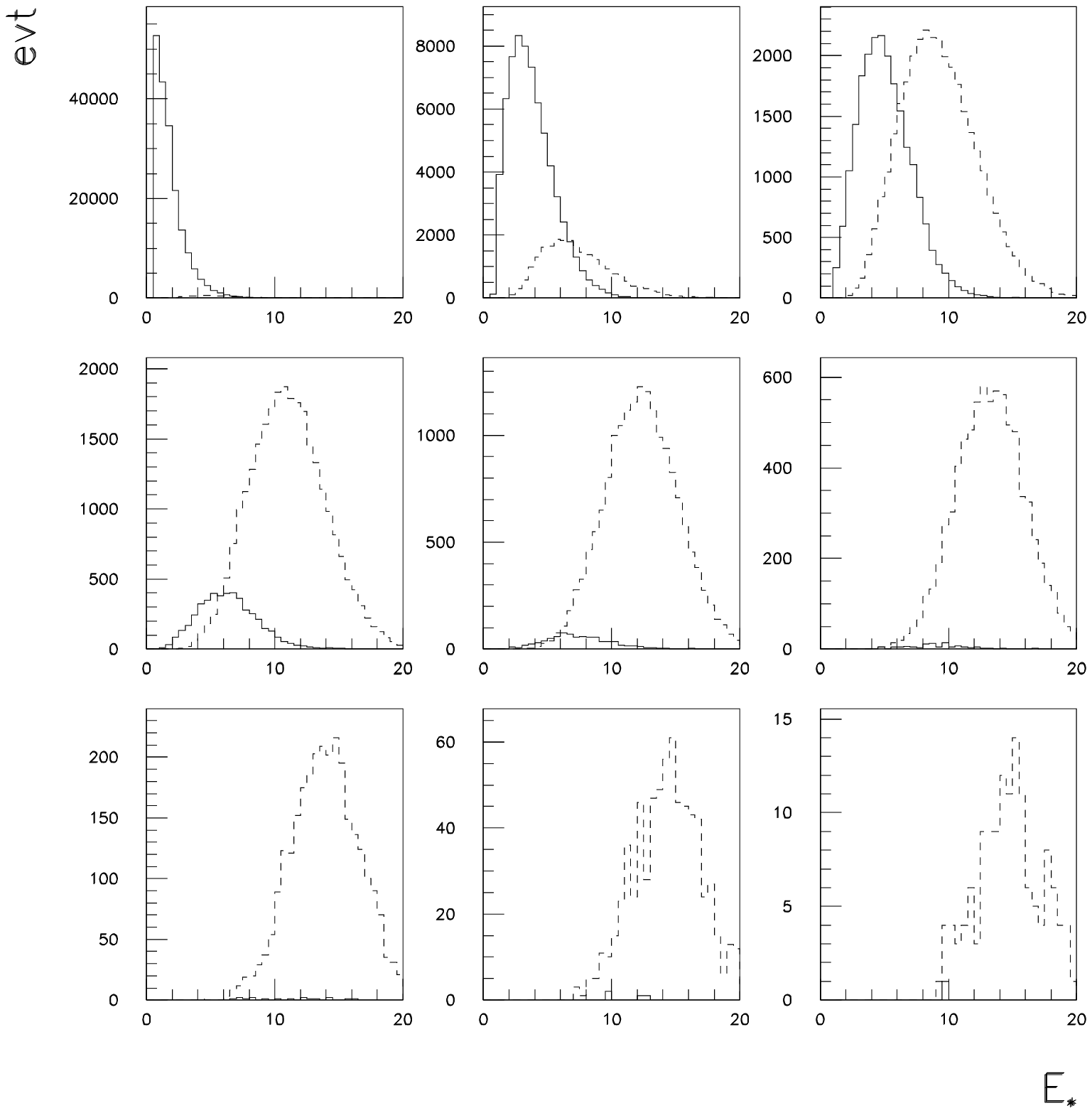}
\end{center}
\vspace{-0cm}
\caption{Excitation energy (MeV/u) distributions for various $Etrans$ bins. The events 
have been separated in two classes corresponding to the 2 bimodality
solutions: $Z_{asym} \geq 0.7$ (full line histograms) and $Z_{asym} \leq 0.3$ (dashed line histograms). 
Xe+Sn system at 80 MeV/u.}
\label{fig:excit_ener}
\end{minipage}
\end{figure}

To progress in the understanding of the bimodality signal, in this section, we try to connect its properties to 
signals of liquid-gas transition. If fragmentation can be associated to such a transition, then a 
bimodality should be observed in the probability distribution of the system density which is the order parameter 
of the transition. Two measurable variables strongly 
correlated to the density are the size of the heaviest fragment and the energy: both should have bimodal 
probability distributions in the liquid-gas
transition as illustrated in the framework of the lattice-gas model in figure \ref{fig:gulmi_correl}. 
On the experimental side, the matter density is not accessible but the excitation 
energy can be obtained from calorimetry as described in section \ref{QPproperties}.

\subsection{QP reconstruction}\label{QPproperties}
In order to be able to characterize the QP, one has to identify its centre of mass and its excitation energy. 
As discussed in section \ref{sorting},
any product emitted forward in the ellipsoid frame has been attributed to the QP. 
In order to minimize the contribution of midrapidity, 
we have calculated the origin of the QP frame from the IMF only (Z $\geq$ 3) at variance from 
section~\ref{mid-rapid-substr}. Then, we have estimated the deposited energy from
calorimetry in including all the IMF 
and only the LCP forward emitted in the QP frame. The LCP contribution has been doubled to account for 
backward emitted LCP. This procedure permits to avoid to a large extent LCPs resulting from mid-rapidity emission. 
The charge of the original QP has been determined by adding the detected product charges. Its mass has been 
taken as that giving the same N/Z ratio as the projectile. The masses of the fragments (for Z exceeding 4) 
have been given by the EPAX formula\cite{Summ} and the neutron number has then been obtained from the difference between 
the initial QP mass and the total masses attributed to the detected products. Then the excitation energy has been 
calculated from:
\begin{equation}
  E^{*}=\sum_{i=1}^{M_{IMF}}T_{i}+\sum_{j=1}^{M_{LCP}}T_{j}+ M_{neut} \langle T_{n} \rangle -Q
  \label{calorimetrie}
\end{equation}
where the $T_{i,j,k}$ values are the kinetic energies for each product, $M_{IMF}$, $M_{LCP}$ and $M_{neut}$ are the 
multiplicities of IMF, LCP and neutrons and $Q$ the mass balance. The average kinetic energy of neutrons 
$\langle T_{n} \rangle$ has been obtained by means of an effective temperature\cite{Hauger,Dag02}.

\subsection{Excitation energy distributions}
In figure \ref{fig:excit_ener},
the excitation energy distributions are shown for various $Etrans$ bins and for the Xe + Sn system at 80 MeV/u. 
For each $Etrans$ bin, we have separated the events corresponding to $Z_{asym}$ larger (resp. lower) than
0.7 (resp. 0.3). The two distributions have similar amplitudes in the third bin in which the 
bimodality signal is observed (see figure \ref{fig:Zasym-zmax-Xe}) but they do not coincide. Even if excitation energies
as high as 15 MeV/u are certainly not realistic, we can see that 
a larger mean excitation energy is measured for the multifragmentation case (low $Z_{asym}$) in qualitative agreement with 
the calculation results of figure \ref{fig:gulmi_correl}. Many reasons can be invoked to explain the fact that 
these two distributions overlap 
more strongly than the model calculation results: the sorting is not a canonical one; moreover,
the calorimetry method is far from being perfect 
due to the doubling procedure and to the estimated neutron correction\cite{Vient}.

\begin{figure}[htb]
\begin{center}
\includegraphics[width=9cm]{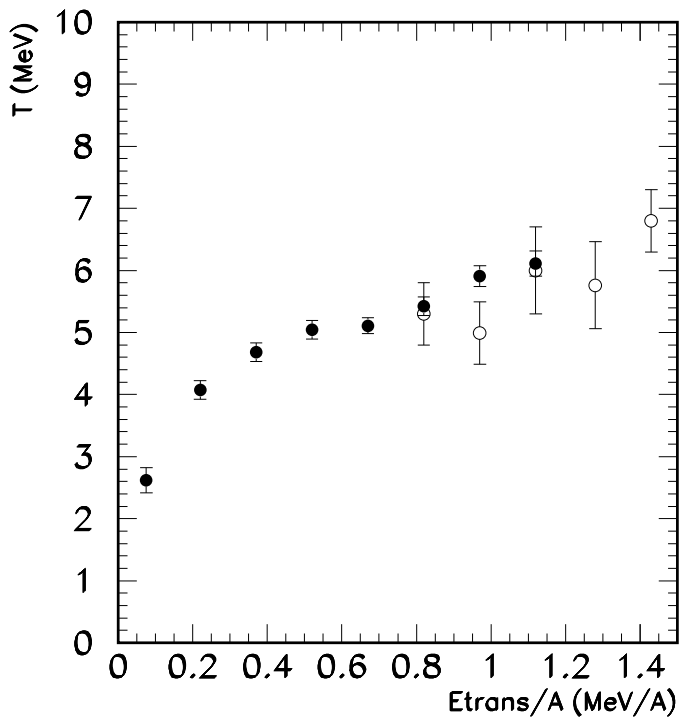}
\caption{Temperatures measured as a function of $Etrans$ for residue and 
multifragment events. Xe+Sn system at 80 MeV/u. The black (resp. open) 
points correspond to two thermometers: slope parameter of protons 
(resp. double isotopic ratio method). The error bars are statistical. The transition region is around
$Etrans/A=1 MeV/u$ (figure \ref{fig:sautAu}).} 
\label{fig:temp}
\end{center}
\end{figure}

\subsection{Temperatures}
One may now try to measure the involved temperatures. If bimodality has a thermal origin,
one would expect to observe similar temperatures for both classes of events: residue-like events and multifragmentation.

Temperature measurement is a difficult task because the various possible thermometers are 
not equally suited to the different energy regimes. The kinetic energy slope method is expected to be adapted when a 
residue is released, but not in the multifragmentation case because the particle
source is not unique. The double isotopic ratio is better suited in this second case 
because the system can explore a small density region\cite{Tsang} and because the IMF multiplicity is significant.
In figure \ref{fig:temp}, we indicate the corresponding results for the four first bins of figure \ref{fig:excit_ener}. 
The abscissa is the transverse energy normalized to the system size and 
the ordinate indicates the measured temperatures. The double isotopic ratio method (He-Li nuclei) 
has been used for the events 
selected with $Z_{asym}$ lower than $0.3$ and the corresponding results have been corrected for side feeding effects\cite{Tsang};
the slope of proton spectra (forward emitted in the QP frame)
has been used for the residue events 
($Z_{asym}$ larger than $0.7$).
The points with low statistics have not been plotted. In the transition energy region ($0.8-1.2$ MeV/u), both groups
exhibit similar temperature values as expected if bimodality has a thermal origin. 

The events belonging to the two regions of the bimodal distribution are thus found to correspond to different excitation
energies but to similar temperatures. It is what is expected for a first order phase transition for 
which the energy jump between the two solutions reflects the latent heat. 
This result is hence coherent with a thermal origin of multifragmentation but cannot be considered as a strong 
evidence because the compared temperatures have been obtained with two different thermometers.

\section{Bimodality and other possible phase transition signals}
As mentioned in the introduction, several signatures of phase transition have
been proposed and experimentally observed. Due to the various experimental
conditions, each of these signals present some weaknesses. Thus the occurrence
of a phase transition would be strongly reinforced by a simultaneous 
observation of several signatures. Such is the aim of this section.

\begin{figure}[htb]
\begin{minipage}[t]{65mm}
\begin{center}
\includegraphics[width=6.5cm]{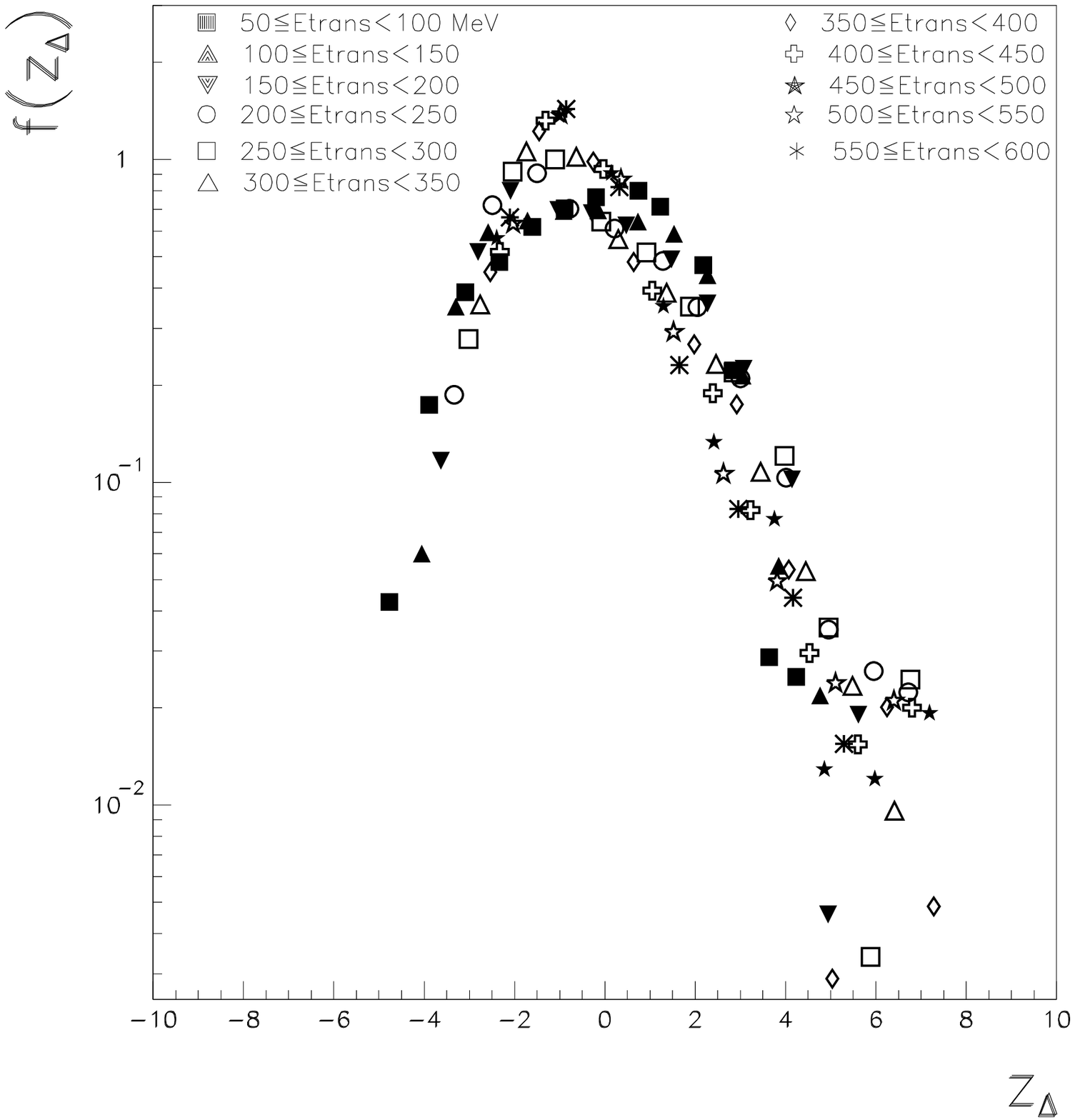}
\end{center}
\vspace{-0cm}
\caption{Normalized $Z_{max}$ distributions according to equation \ref{eq2} for the system Xe+Sn at 80 MeV/u 
with $\Delta $=0.5. Various symbols correspond 
to various $Etrans$ bins.} 
\label{fig:deltascaling1}
\end{minipage}
\hspace{\fill}
\begin{minipage}[t]{65mm}
\begin{center}
\includegraphics[width=6.5cm]{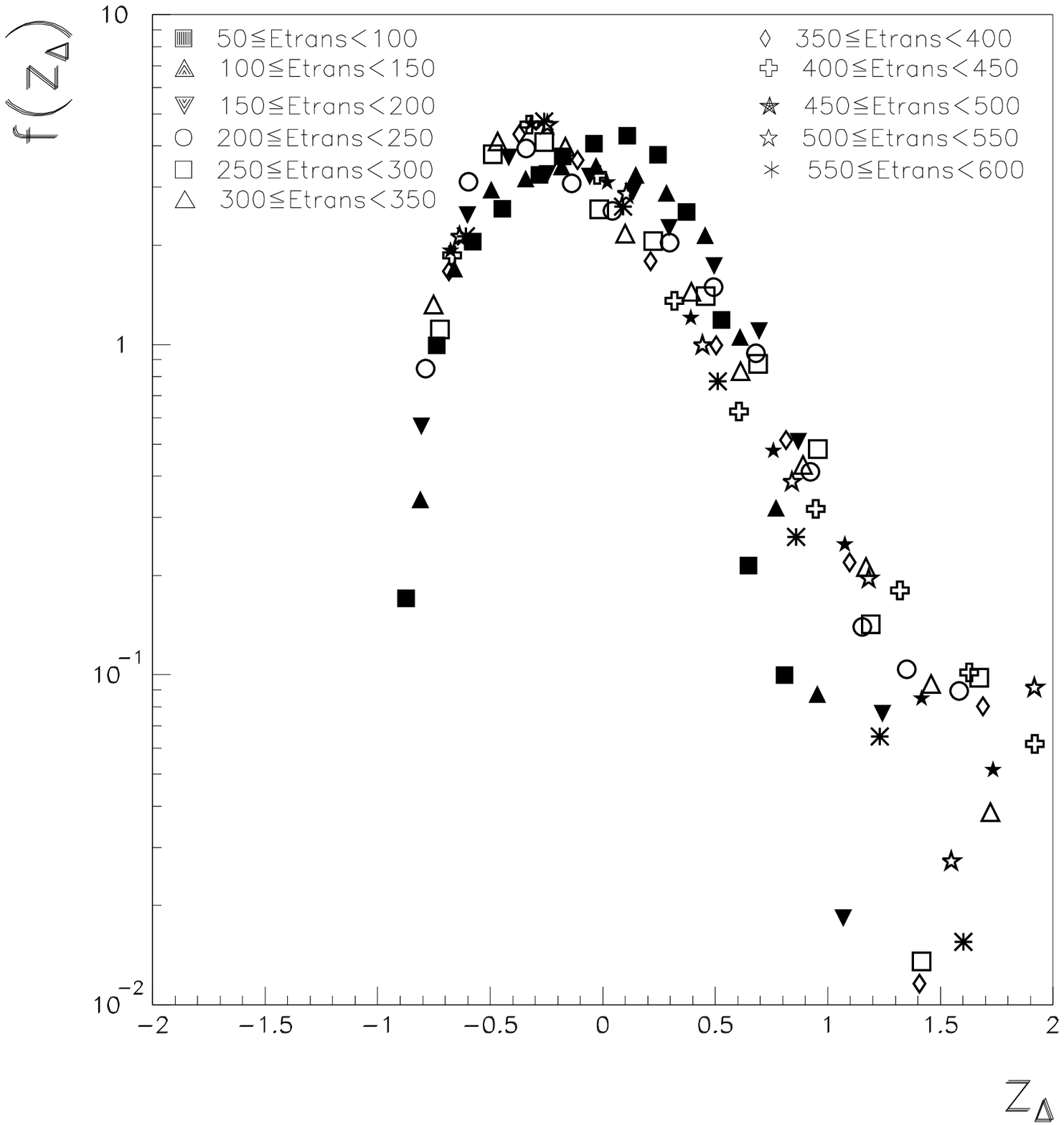}
\end{center}
\vspace{-0cm}
\caption{Similar to figure \ref{fig:deltascaling1} with $\Delta $=1.}
\label{fig:deltascaling2}
\end{minipage}
\end{figure}

\begin{figure}[htb]
\begin{minipage}[t]{65mm}
\begin{center}
\includegraphics[width=6.5cm]{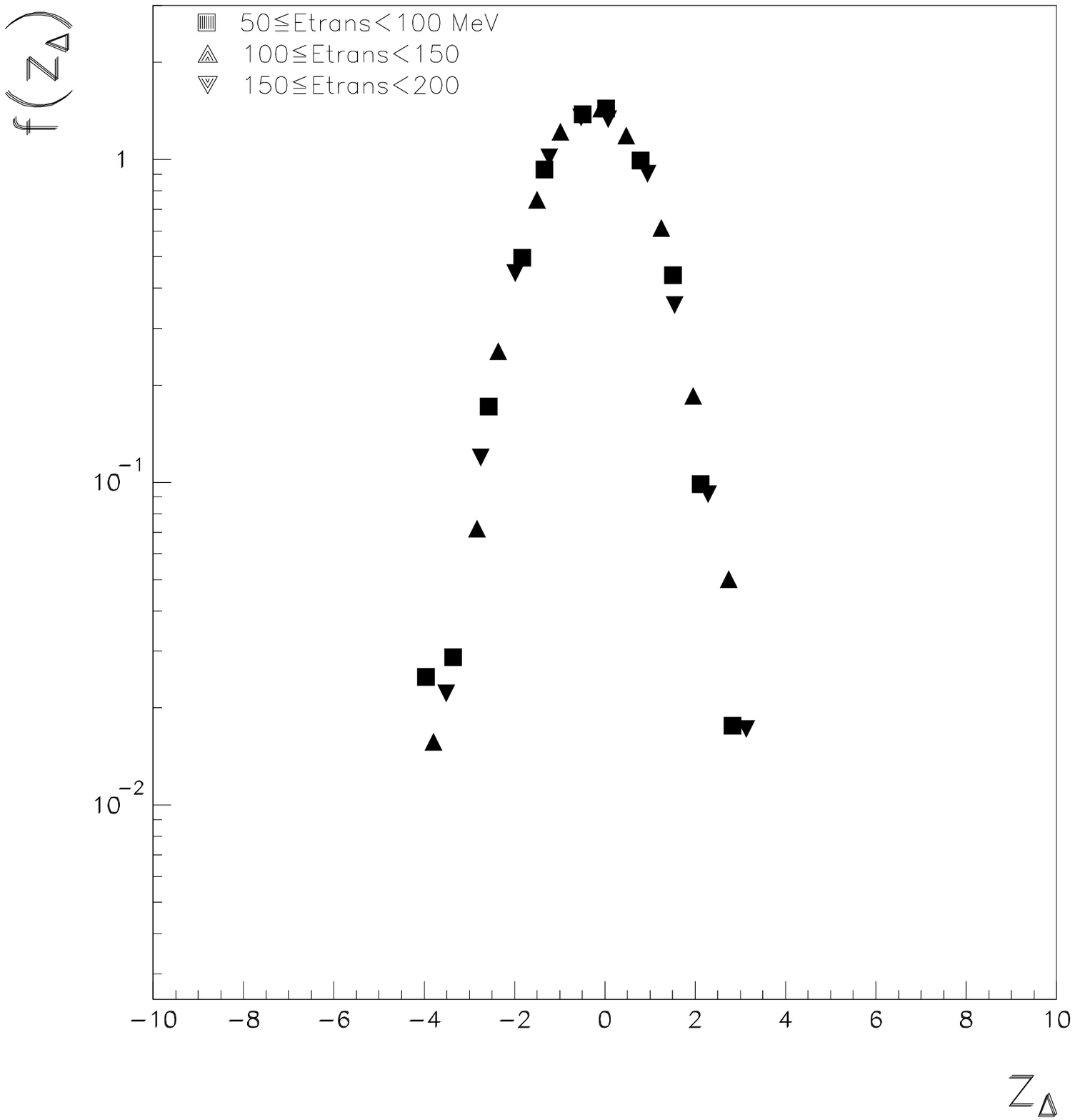}
\end{center}
\vspace{-0cm}
\caption{Delta scaling for the system Xe+Sn at 80 MeV/u with $\Delta $=0.5. 
Only residue like events ($Z_{asym}$ larger than 0.7) have been retained. 
Various symbols correspond to various $Etrans$ bins. Only the $Etrans$ bins for which 
the residue contribution is dominant have  been retained.}
\label{fig:deltascaling-tries1}
\end{minipage}
\hspace{\fill}
\begin{minipage}[t]{65mm}
\begin{center}
\includegraphics[width=6.5cm]{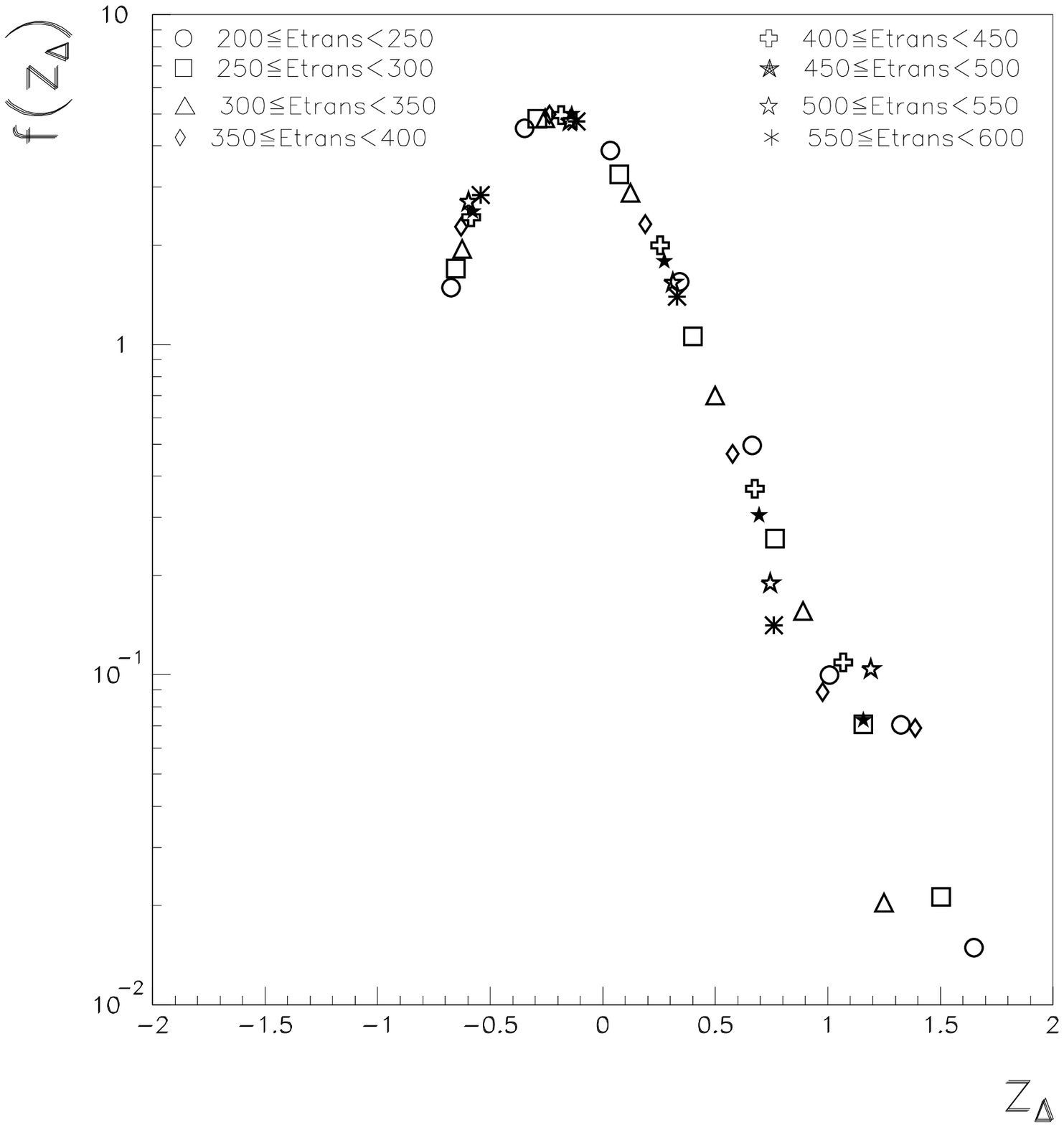}
\end{center}
\vspace{-0cm}
\caption{Delta scaling for the system Xe+Sn at 80 MeV/u with $\Delta $=1. 
Only multifragmentation events ($Z_{asym}$ lower than 0.3) have been retained. 
Various symbols correspond to various $Etrans$ bins. Only the $Etrans$ bins for which 
the multifragmentation contribution is dominant have  been retained.}
\label{fig:deltascaling-tries2}
\end{minipage}
\end{figure}

\subsection{Relation between bimodality and $\Delta$-scaling}

$\Delta$-scaling is a signal proposed in the framework of the universal fluctuation theory \cite{Botet}.
It may be used to distinguish between different phases and to identify critical points. 
$\Delta$-scaling is observed for an order parameter $m$ if the probability $P_{N}(m)$ for observing a given 
value of $m$ can be written with the following formula 
\begin{equation}
<m>^{\Delta }P_{N}(m)= f(z_{\Delta })= f(\frac {(m-<m>)}{<m>^\Delta })
 \label{eq2}
\end{equation}
in which $f(z_{\Delta })$ is a universal 
function which does not depend on the size $N$ of the system. For a system exhibiting two phases,
the $\Delta$ value is expected to be close to $0.5$ for an ordered phase and to $1$ for a disordered 
phase. The transition point between the two phases is first order if, at this point, the distribution of $m$ is bimodal, 
and second order if it obeys $\Delta$=1 scaling with a non-Gaussian tail\cite{Botet}. Several data have been analyzed 
in this context\cite{Botet2,Frankland}. In reference \cite{Frankland} $\Delta$-scaling was evidenced for central 
collisions for the Xe + Sn and Au + Au systems of interest in the present paper. Those results indicate that the size 
of the largest fragment is an order parameter. 
The ordered (resp. disordered) behaviour is observed below (resp. above) an incident energy 
of 39 MeV/u for the Xe+Sn system. In these works, the nature of the transition could not be determined. 

We tested such a behaviour for the semi-peripheral collisions of interest in the present paper.
In figures \ref{fig:deltascaling1} and \ref{fig:deltascaling2} the $f(z_{\Delta })$ functions (with $m=Z_{max}$)
are plotted for various $Etrans$ bins 
for the 80 MeV/u Xe + Sn system. The plots correspond to $\Delta$  values of 0.5 and 1 respectively. 
No scaling is observed. However, scaling is recovered if one separates the events according to the 
$Z_{asym}$ ranges corresponding to defined events: residues ($Z_{asym}$ larger than 0.7)
or multifragmentation  ($Z_{asym}$ smaller than 0.3).  
The results are shown in figures \ref{fig:deltascaling-tries1} and \ref{fig:deltascaling-tries2}. 
Residue events are numerous only 
for $Etrans$  values smaller than 200 MeV. Only the corresponding bins have been retained in figure 
\ref{fig:deltascaling-tries1}. 
Conversely, multifragmentation becomes dominant for $Etrans$ larger than 200 MeV. Only the corresponding bins have been 
retained in figure \ref{fig:deltascaling-tries2}. Similar results have been obtained for the Au+Au system with 
fission events reconstructed as explained in section 3\cite{Pichon2}.

The fact that the scaling is clearly observed for the two classes of events suggests that these latter can be understood
as an ordered (resp. disordered) phase. Conversely, the absence of scaling for the global distribution is consistent 
with the interpretation in terms of phase coexistence. Indeed, the absence of bimodality in the
distribution of $Z_{max}$ may be due to a too strong correlation between the sorting variable and the deposited 
energy, which is also an order parameter: if a constraint is applied on an order parameter in the coexistence zone, its
bimodal character is naturally suppressed. As the system is still in the coexistence zone, 
a pure phase scaling may be expected to fail as it is suggested in recent theoretical calculations\cite{Gulmi2005}.

\subsection{Relation between bimodality and negative heat capacity}\label{negative}
Another signal of phase transition is negative heat capacity. This phenomenon is univocally associated to first order phase 
transitions with a finite latent heat in isolated systems with 
a size comparable to the range of the forces governing its equation of state\cite{Gross}. 
A negative heat capacity occurs whenever the entropy presents
an anomalous curvature as a function of the available energy. It has been shown 
in reference \cite{Chomaz99} that
heat capacity may be deduced from the measurement of the fluctuations in the sharing 
of the available energy between independent degrees of freedom. 
Since the heat capacity can be negative only in the microcanonical ensemble, from an experimental point of view 
events with a defined excitation energy have to be selected. 
For such a sorting, the total heat capacity may be approximately written:
\begin{equation}
C \approx \frac {C_{1}^{2}}{C_{1}-\frac {{\sigma} ^{2}}{T^{2}}}
\label{eq3}
\end{equation}
where $C_{1}$ is the partial heat capacity for 
a subset of degrees of freedom (for instance momentum space), $T$ is the temperature and 
${\sigma} ^{2}$ is the variance associated with the sharing of the total available energy between 
the degrees of freedom subset and the other ones. The total heat capacity becomes negative when these fluctuations 
become larger that the fluctuations associated to the canonical ensemble.

This formalism has been used to study central Xe + Sn collisions from 32 to 50 MeV/u \cite{Lenein}, and 
peripheral Au + Au collisions at 35 MeV/u \cite{Dagos}. In both cases, fluctuations exceeding the 
canonical expectation were found. For the
Xe + Sn central collisions, the signal is observed at 32 and 39 MeV/u but not at larger bombarding
energies for which the system would have left the coexistence region. 

We have tried to correlate this signal with the bimodality one which is discussed in this paper. For this purpose, 
events must be sorted according to the excitation energy of the QP, obtained by calorimetry on the QP source
as explained in section \ref{QPproperties}. We have retained only events with a total reconstructed source charge 
equal to the projectile charge within 10\%.

The negative heat capacity study has been performed in studying the sharing of the excitation energy between the kinetic 
and the potential energy parts at the freeze-out configuration. 
Rewriting equation \ref{eq3} in normalizing the heat capacity 
to the source size gives: 
\begin{equation}
  C \approx \frac{C_{k}^{2}}{C_{k}-\frac{<A_{sou}>\sigma_{k}^{2}}{T^{2}}}
 \label{Capa}
\end{equation}

$\sigma_{k}^{2}$ is the variance of the kinetic energy distribution; 
$<A_{sou}>$ is the average mass number of the source; 
$C_{k}$ is the partial kinetic energy heat capacity per nucleon: 
\begin{equation}
  C_{k}=\frac{d<\frac{E_{k}}{A_{sou}}>}{dT}
 \label{Ck}
\end{equation}
The potential energy $E_{p}$ has been obtained from the mass balance corrected for the Coulomb part in assuming that all the LCP are 
emitted from the IMF in a second step after the freeze-out:
\begin{equation}
  E_{p}=(\sum_{i=1}^{M_{IMF}}m_{i}-m_{sou})+E_{coul}
\end{equation}
$M_{IMF}$ is the IMF multiplicity ($Z \ge 3$); the $m_{i}$ are the IMF primary masses; $m_{sou}$ is the reconstructed 
QP mass; the Coulomb energy balance at 
freeze-out $E_{coul}$ has been calculated from the Wigner-Seitz formula\cite{Bond}. 

The temperature of relation \ref{Capa} has been calculated from the total kinetic energy by using for
the level density parameter $a_{i}=\frac{A_{i}}{8}$: 
\begin{equation}
  <E_{k}>=<\sum_{i=1}^{M_{IMF}}a_{i}>T^{2}+<\frac{3}{2}(M-1)>T
\end{equation}

\begin{figure}[htb]
\begin{minipage}[t]{65mm}
\begin{center}
\includegraphics[width=6.5cm]{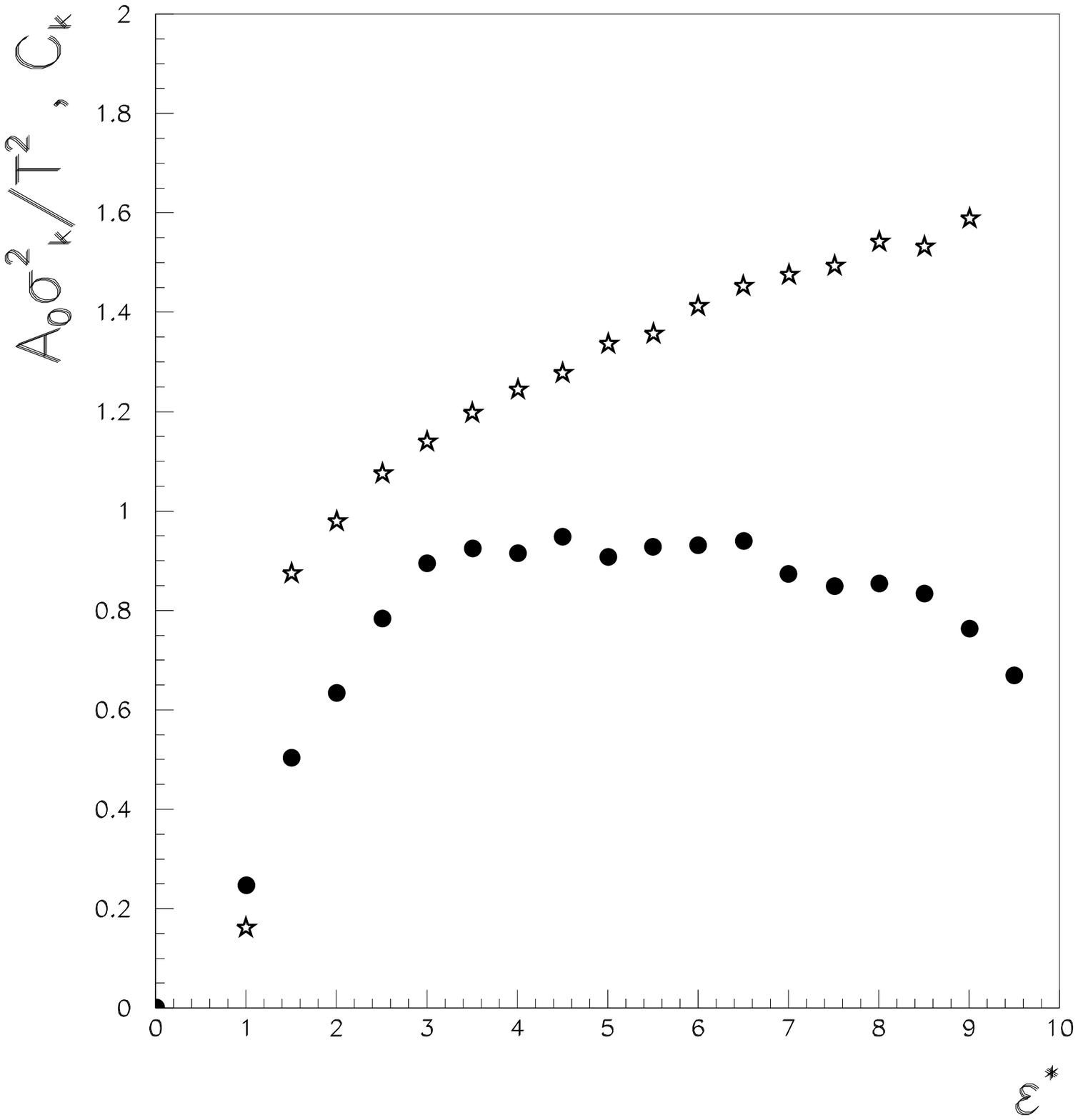}
\end{center}
\vspace{-0cm}
\caption{Evolution with the available excitation energy in $MeV/u$ of the two terms of the denominator
of relation \ref{Capa} (see text): partial heat capacity (open stars), and fluctuation term (black points). 
System Au + Au at 80 MeV/u.}
\label{fig:nega_au80}
\end{minipage}
\hspace{\fill}
\begin{minipage}[t]{65mm}
\begin{center}
\includegraphics[width=6.5cm]{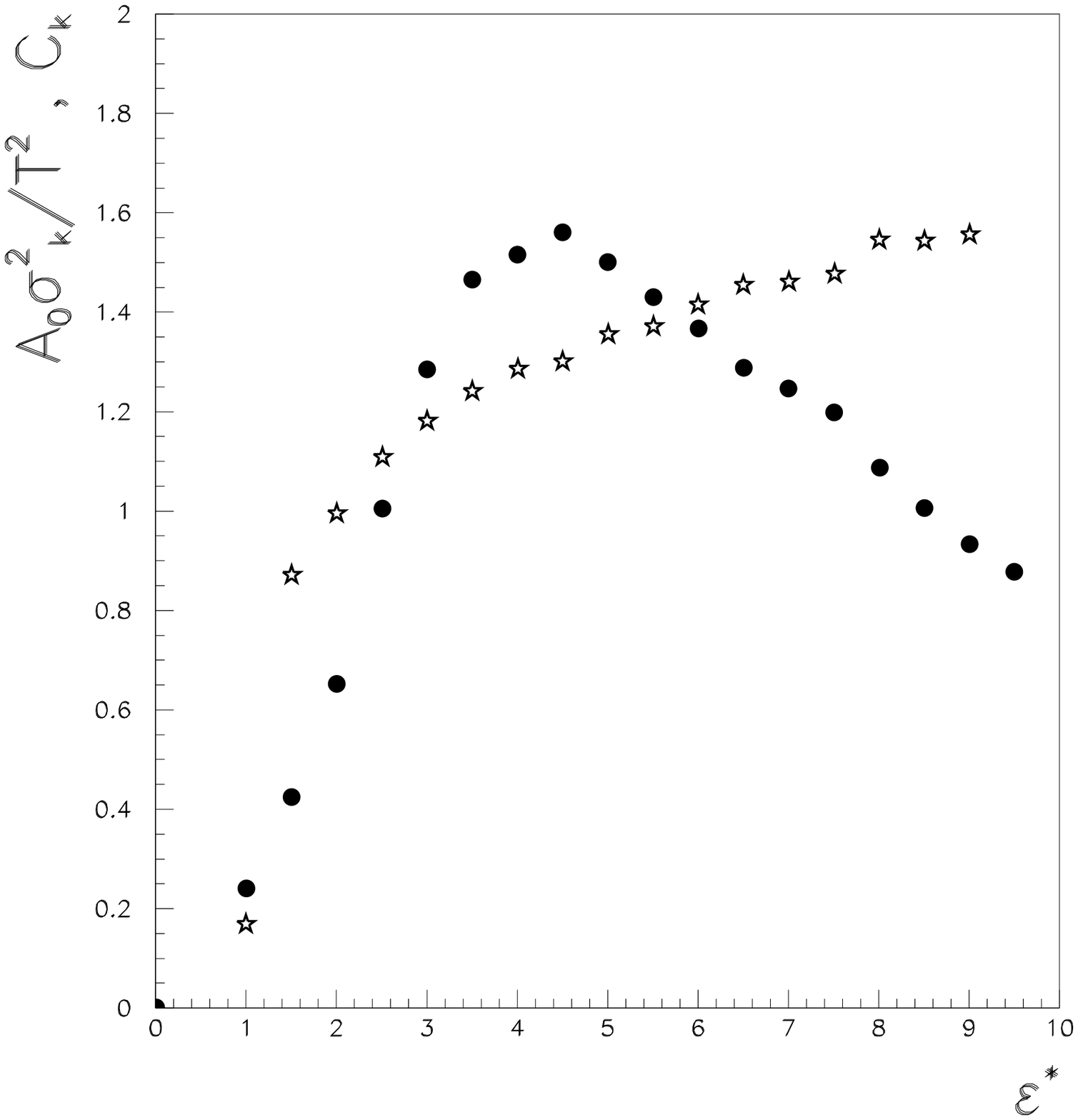}
\end{center}
\vspace{-0cm}
\caption{Similar to figure \ref{fig:nega_au80} but in selecting the most compact QP events: see text.}
\label{fig:nega_au80_compact}
\end{minipage}
\end{figure}

From equation \ref{Capa} it turns out that the heat capacity will be negative 
if the second term of the denominator (fluctuation term) exceeds the first one (partial heat capacity). 
These two terms are plotted in figure \ref{fig:nega_au80}. 
Fluctuations are far below the canonical expectation whatever the excitation energy is. This result is quite 
general and has been observed for quasi-projectiles from the Xe + Sn and Au + Au systems at any bombarding energy.

\begin{figure}[!h]
  \centering
  \includegraphics[height=4.7cm]{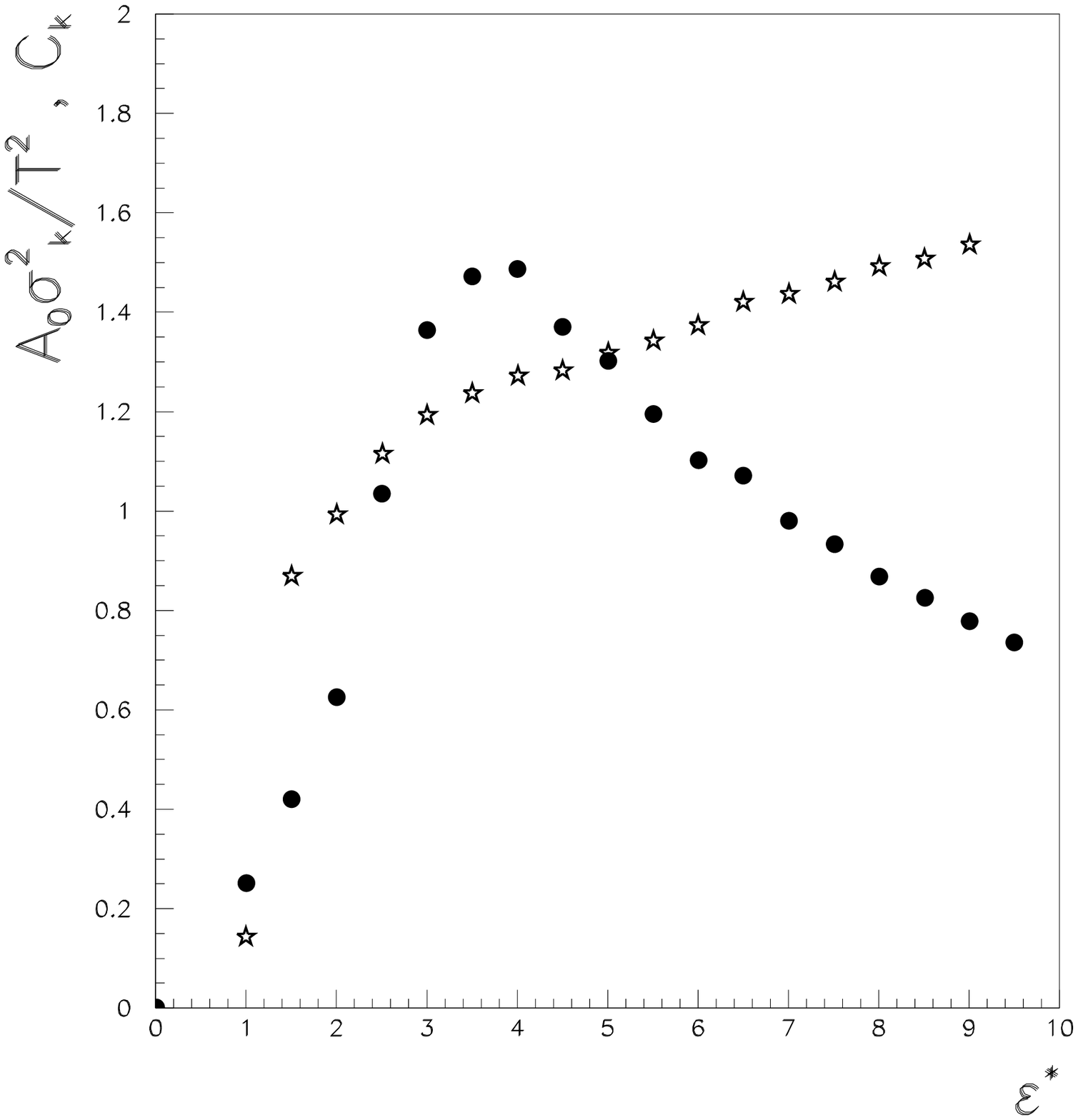}
  \includegraphics[height=4.7cm]{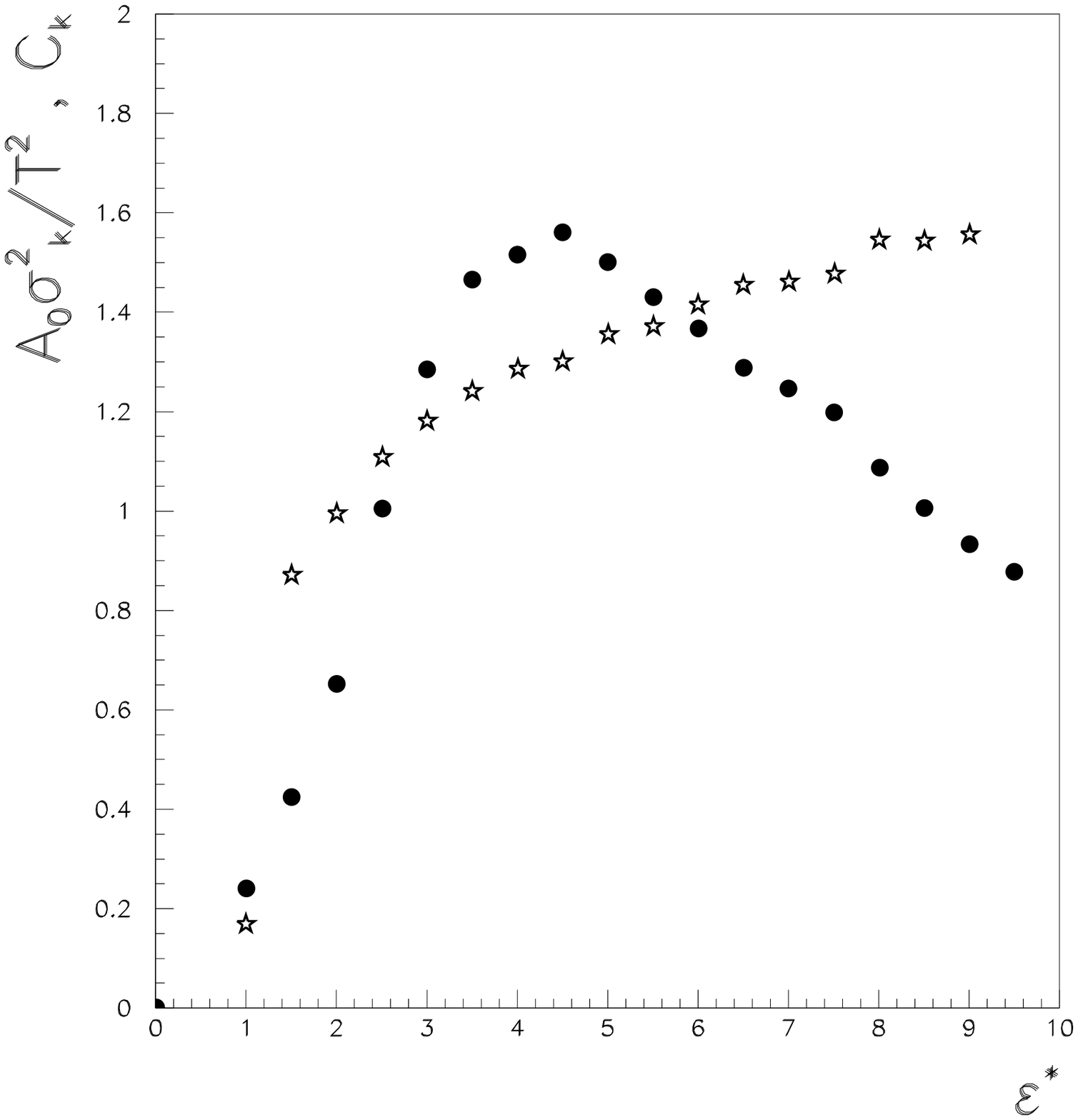}
  \includegraphics[height=4.7cm]{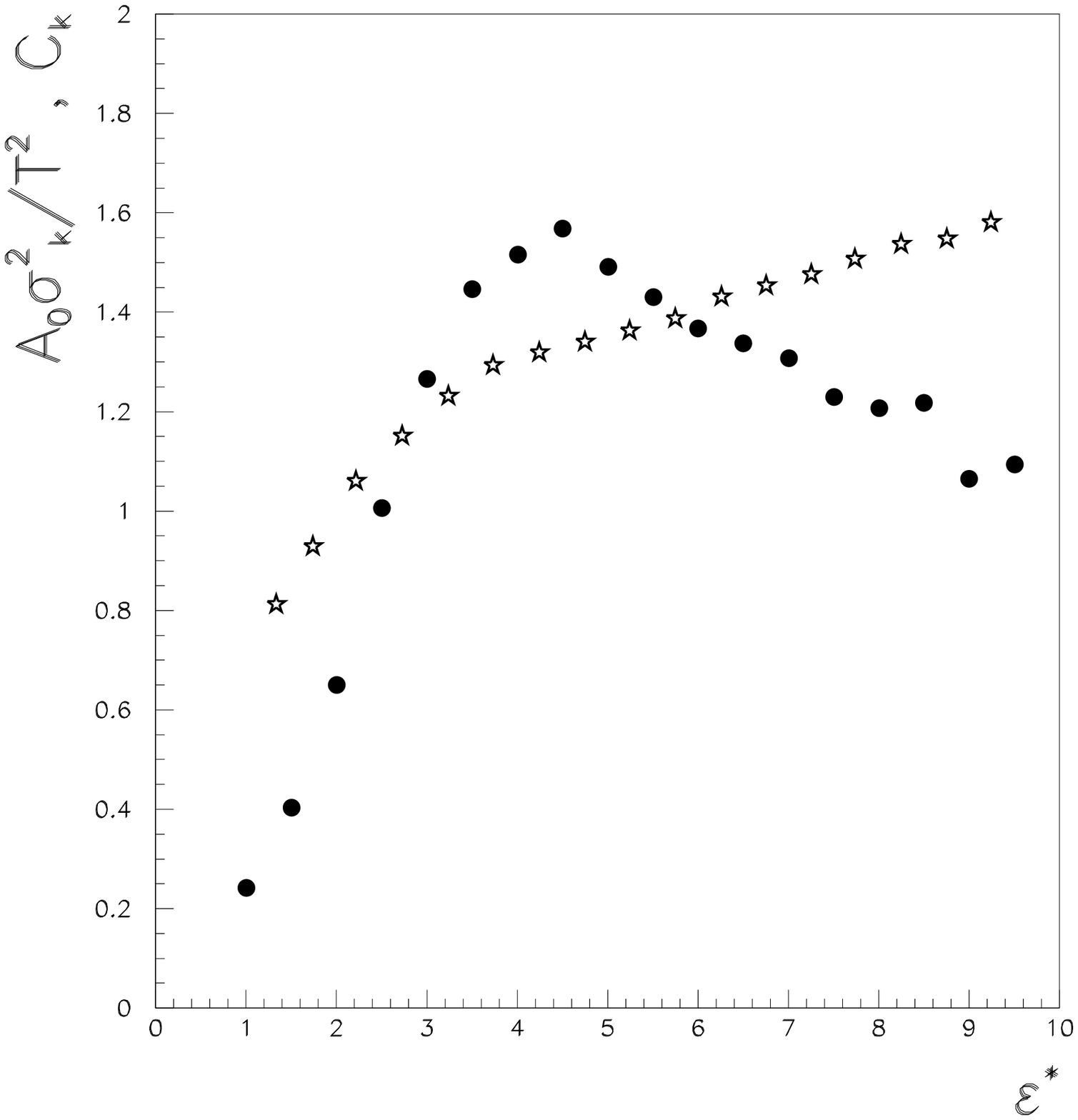}
  \caption{Systems Au + Au at 60 (left), 80 (middle) and 100 (right) MeV/u. 
Evolution with the available excitation energy of the two terms of the denominator
of relation \ref{Capa}: partial heat capacity (open stars), and fluctuation term (black points).
Non compact events (see text) have been  rejected.}
  \label{fig:nega_au_compact}
\end{figure}

\begin{figure}[!h]
  \centering
  \includegraphics[height=4.5cm]{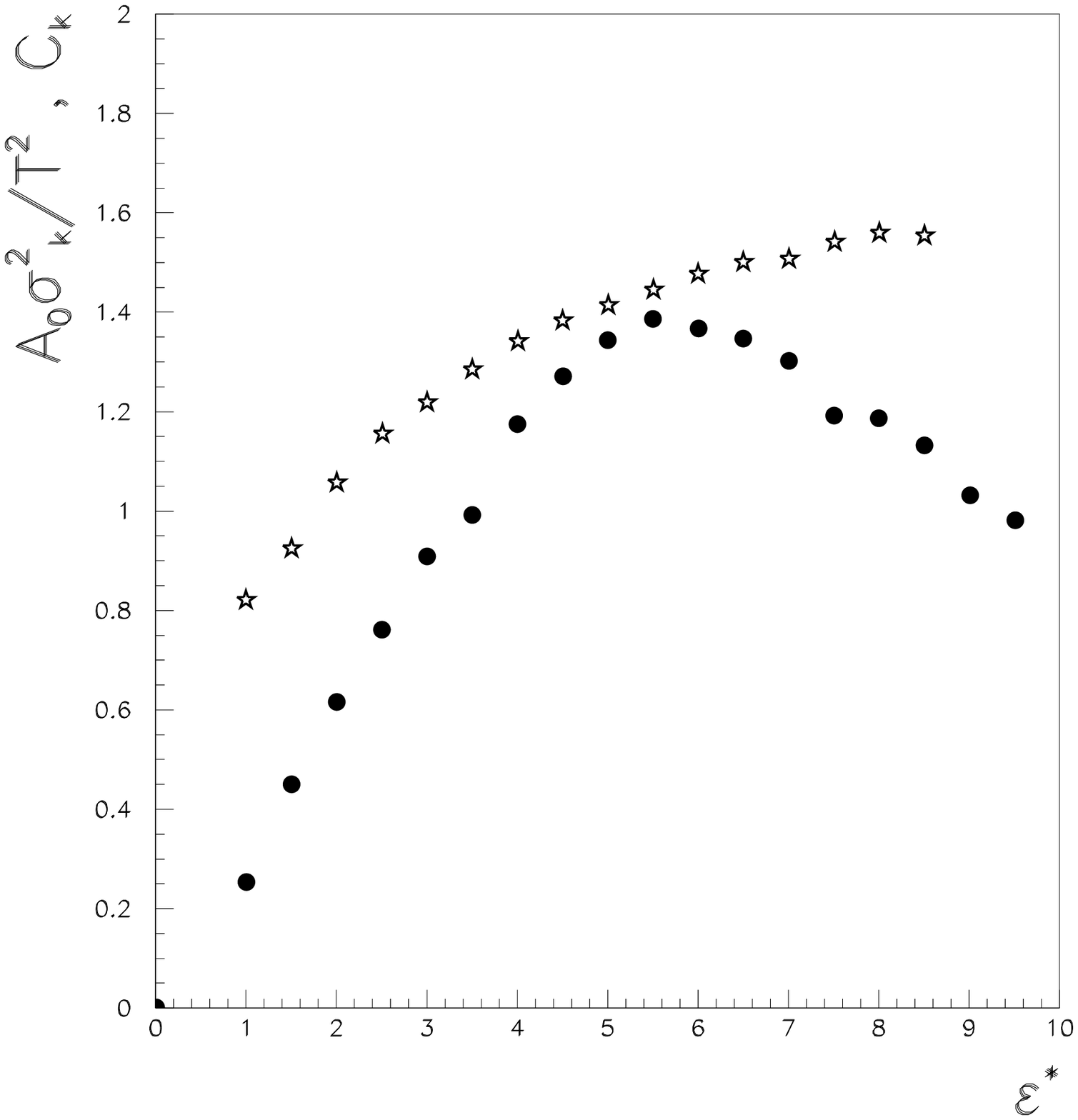}
  \includegraphics[height=4.5cm]{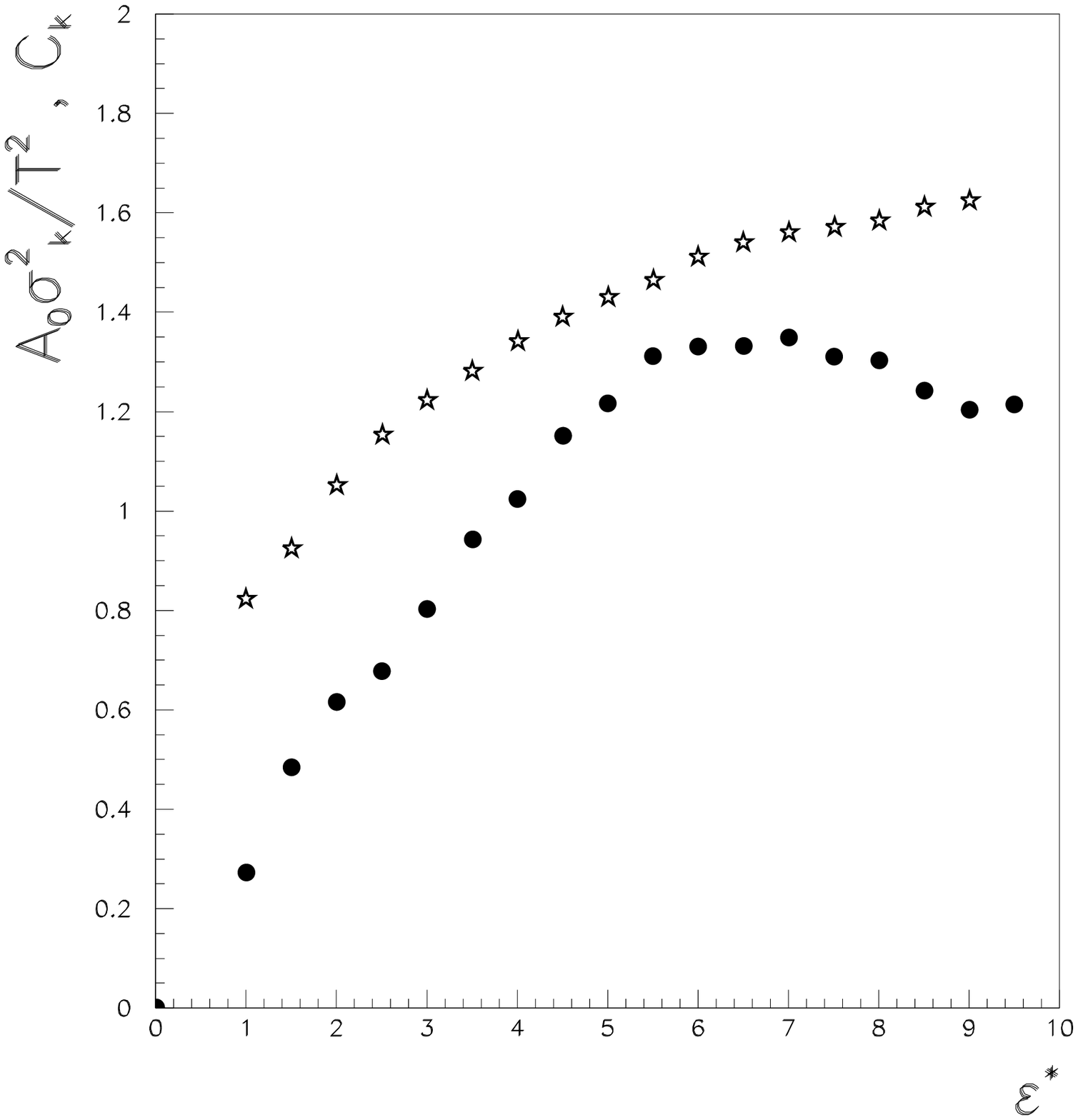}
  \includegraphics[height=4.5cm]{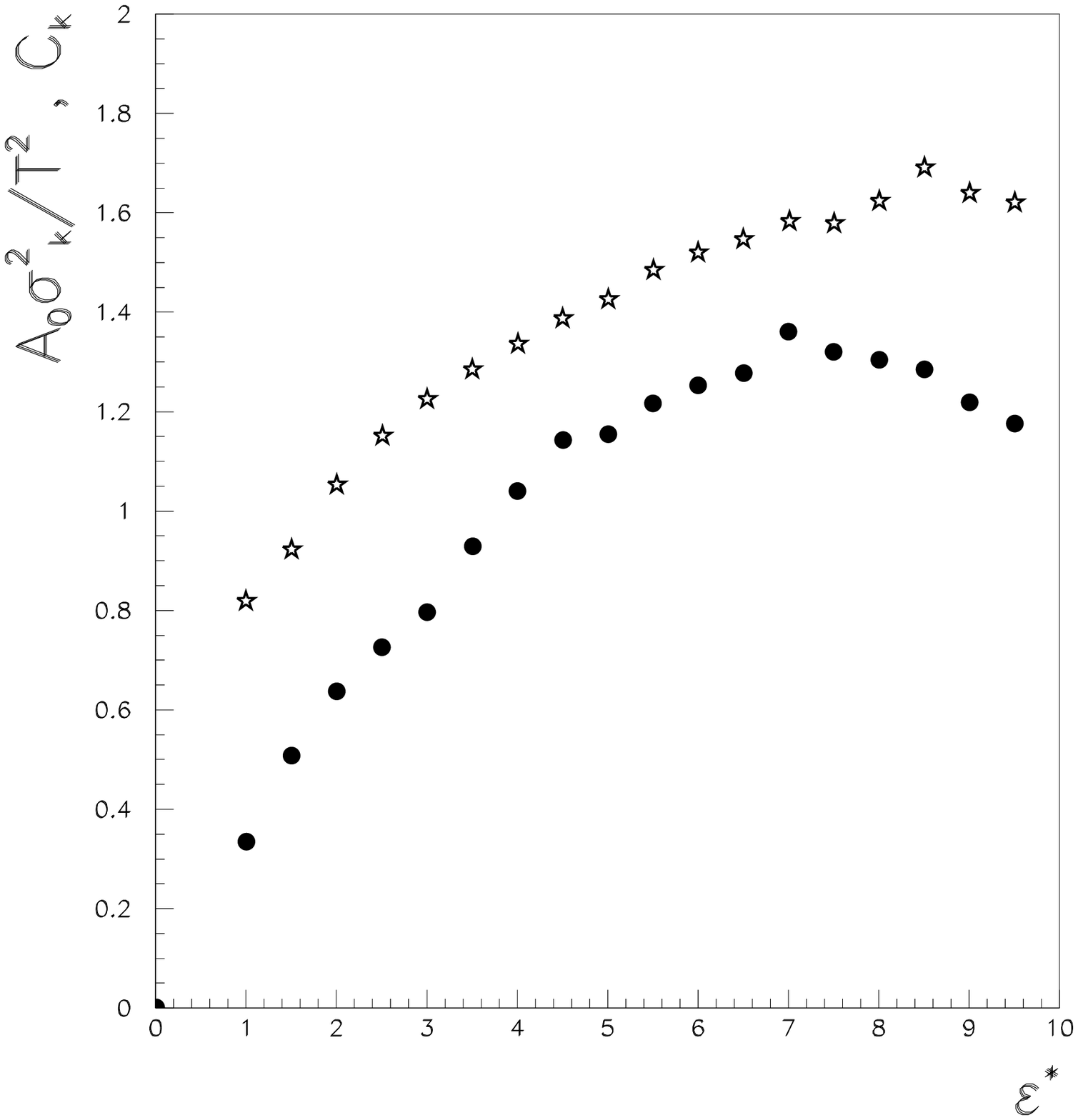}
  \caption{Systems Xe + Sn at 65 (left), 80 (middle) and 100 (right) MeV/u. 
Evolution with the available excitation energy of the two terms of the denominator
of relation \ref{Capa}: partial heat capacity (open stars), and fluctuation term (black points).
Non compact events (see text) have been  rejected.}
  \label{fig:nega_xe_compact}
\end{figure}

At this point, one may recall that the negative heat capacity signal is much more fragile than the bimodality one. 
First, the evaluation of fluctuations is very sensitive to the details of the calorimetric hypotheses 
and side feeding corrections\cite{Palluto}. Moreover, any energy fraction which has not been shared among all the available degrees of freedom but is still stored 
in some selected ones will induce a decrease of the fluctuations we are looking at. In other words, negative heat capacity
is not observed if dynamical effects freeze a sizeable part of the available energy in a few defined degrees of freedom 
reminiscent of the initial direction of the beam. In order to decrease the influence of dynamical effects, we have tried to 
select QP events in which the memory of the entrance channel is minimized. We did such a selection in 
section \ref{dynamic} in cutting the $cos(\theta )$ distribution of figure \ref{fig:disangzmax}). 
This cut does not modify significantly the signal. To ensure a stronger reduction of dynamical effects,
we selected compact events, 
i.e. events for which the QP decay is not elongated in the momentum space. Compactness has been estimated from
the relative velocity between the largest QP fragment and the mean velocity of the other QP fragments:
\begin{equation}
  v_{rel}=v_{Z_{max}}-\frac{\sum_{i=1}^{M_{IMF}-1}A_{i}v_{i}}{\sum_{i=1}^{M_{IMF}-1}A_{i}}
\end{equation}
A larger value $v_{rel}$ is expected if intermediate velocity IMF contribute to the QP. We have retained only the 
events for which $v_{rel}$ was smaller than 3 cm/ns which is a value expected from Coulomb repulsion. The corresponding results 
are shown in figure \ref{fig:nega_au80_compact}. Only with this selection, abnormal fluctuations appear for excitation
energies between $3$ and $5-6$ MeV/u. 
This is a general feature for the Au + Au system as shown in figure \ref{fig:nega_au_compact} while, at variance, 
the fluctuations
remain below their canonical expectation for any excitation energy for the Xe + Sn system 
(see figure \ref{fig:nega_xe_compact}).
This difference between the two systems is not fully understood.

\begin{figure}[htb]
\begin{center}
\includegraphics[width=9cm]{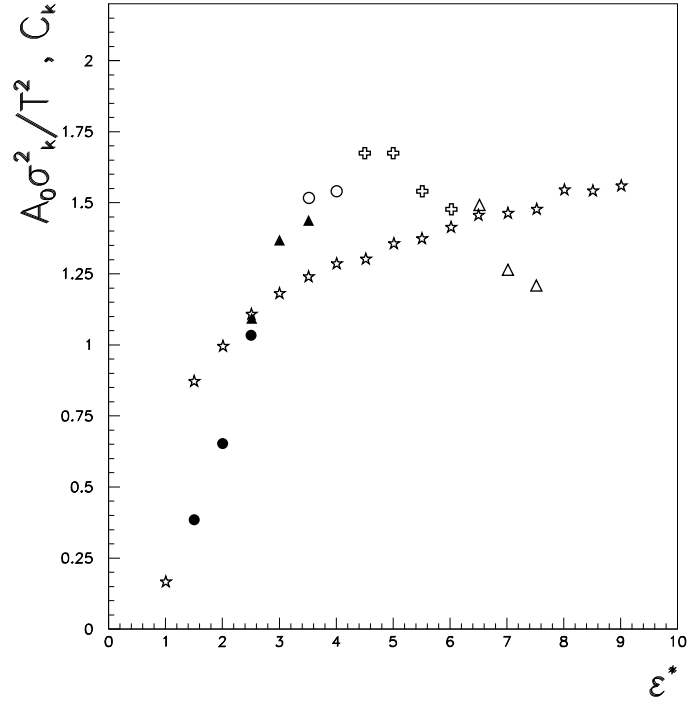}
\caption{Similar to figure \ref{fig:nega_au80_compact} for the Au + Au 80 MeV/u system but the events have been 
sorted according to $Etrans$. Each symbol corresponds to a selected $Etrans$ bin of 0.25 MeV/u width. 
The fluctuation peak is observed in the bins associated with the open points and to the crosses covering 
an $Etrans/u$ range from 0.65 to 1.15 MeV/u) in which the
bimodality transition has been observed in figure \ref{fig:sautAu}.}
\label{fig:nega_au80_etrans}
\end{center}
\end{figure}

Let us focus now on the Au + Au system for which fluctuations are more important. 
One may first remark that the excitation energy range corresponding to the fluctuation peak
is quite independant of the bombarding energy. In figure 
\ref{fig:nega_au80_etrans} we have connected this signal with the bimodality one. Events have been sorted according to 
the $Etrans$ variable 
used previously for bimodality studies and only the most probable excitation energy bins have been retained for each
$Etrans$. Each symbol of figure \ref{fig:nega_au80_etrans} corresponds to an $Etrans$ 
bin. The fluctuation peak corresponds to the third (open points) and fourth 
(crosses) bins, i.e. for $Etrans/u$ values covering the range [0.65-1.15 MeV/u] in which the
bimodality transition has been observed in figure \ref{fig:sautAu}. This coherence indicates that 
abnormal fluctuations and bimodality may sign the same physical phenomenon for the same set of events, but 
sorted in two different ways.

\section{Conclusion}
We studied two symmetrical systems Xe + Sn and Au + Au from 60 to 100 MeV/u. We focused on semiperipheral
collisions for which it is possible to isolate a QP contribution. The sorting of the collisions has been achieved 
from the LCP transverse energy on the QT side and we observed QP properties thus avoiding as much as possible 
auto-correlation due to sorting. We selected events for which most of the available charge on the QP side has been detected 
and identified. The distribution of the charge of the heaviest fragment and the asymmetry between the two heaviest
fragments produced in each event was analysed. We have observed a two-peaked distribution for all systems
for a temperature range around 5 MeV. One peak corresponds to events dominated by a residue and the other one 
to multifragmentation.
Fission events in the Au case have been treated together with the residue-like events. 
If this behaviour is strongly suggestive of a phase transition, it does not allow to conclude about the nature 
of the phenomenon. 

We observed a scaling on the total mass 
of the system since the transverse energy for which bimodality is observed is proportional to this mass for a 
defined bombarding energy per nucleon. On the other hand, we observed a transition transverse energy roughly 
proportional to the beam energy. We interpreted this effect as due to pre-equilibrium effects which increase 
with the bombarding energy. This statement is supported by the fact that the bimodality signal is observed for 
smaller transverse energy values if one selects events which are closer to equilibrium. This interpretation
is still open to discussion. Nevertheless, it turns out that many features are in agreement with an 
interpretation in terms of a phase transition in which the major parameter is the deposited energy. As a matter of 
fact, it has been established that the deposited (excitation) energies at the transition are larger for multifragmentation than for
the residue-like events. At the same time, it has been shown that the temperatures involved in both cases are similar. 
This second conclusion is however weakened by the fact that the available thermometers are different for the two sets 
of events. 

A very interesting feature is that the bimodality signal can be correlated with defined scaling properties 
in the distribution of the heaviest fragment size and the observation of important configurational energy fluctuations
which we have tentatively associated at least for the Au+Au system to negative heat capacity. 

Altogether this set of results is in agreement with the expectations from a phase transition belonging to the liquid-gas 
universality class, and is well reproduced by a dynamical model. Many points need to be clarified. If the convexity 
can be seen in the $Z_{max}$-$Z_{asym}$ representation, this convexity is not very pronounced and cannot be observed 
in the $Z_{max}$ distribution. This leaves the question of the order of the transition open. The relevant order parameter
and the associated nature of the transition is not clear either. In particular the dynamical code HIPSE suggests that 
angular momentum may play a major role in the transition. 

Up to now, bimodality has been mainly recognized in semiperipheral collisions. The fact that it is more difficult 
to observe it in central ones may reflect a lower involved angular momentum. But it can also be due to the fact that 
the central collision sorting is too close to a microcanonical selection (for which bimodality should not 
be observed) rather than a canonical one ensuring the energy fluctuations which are necessary to observe 
the coexisting phases. 
In order to better elucidate this question, we intend to progress in two directions. We will first concentrate on a
 better understanding of central collisions in mixing various bombarding energies to simulate a canonical sorting. Another kind
of analysis will concern the understanding of the role of the distance to equilibrium which is reached for a set of events. 
We will address this question in comparing collisions with an entrance channel which is asymmetric in isospin. 
The distance to equilibrium will be inferred from the distance to equilibrium in the isospin degree of freedom. 
Such measurements have  been performed with INDRA at GSI for the $^{129}Xe + ^{112}Sn$ et $^{124}Xe + ^{124}Sn$ at 
100 MeV/u.

\end{document}